\documentclass[12pt]{article}

\usepackage{amsmath}
\usepackage{amssymb}
\usepackage{amsfonts}
\usepackage{latexsym}
\usepackage{color}
\usepackage{subfigure}
\usepackage{graphicx,graphics,epstopdf}
\catcode `\@=11 \@addtoreset{equation}{section}
\def\theequation{\arabic{section}.\arabic{equation}}
\catcode `\@=12

\newcommand{\be}{\begin{equation}}
\newcommand{\en}{\end{equation}}
\newcommand{\bea}{\begin{eqnarray}}
\newcommand{\ena}{\end{eqnarray}}
\newcommand{\beano}{\begin{eqnarray*}}
\newcommand{\enano}{\end{eqnarray*}}
\newcommand{\bee}{\begin{enumerate}}
\newcommand{\ene}{\end{enumerate}}

\newcommand{\mc}{\mathcal}

\newcommand{\D}{{\mc D}}

\newcommand{\Sc}{{\cal S}}

\newcommand{\F}{{\cal F}}
\newcommand{\G}{{\cal G}}

\newcommand{\Lc}{{\cal L}}
\newcommand{\ltwo}{{\Lc^2(\mathbb{R})}}
\newcommand{\scr}{{\cal S}(\mathbb{R})}

\newcommand{\1}{1 \!\! 1}

\newcommand{\Hil}{\mc H}

\newtheorem{thm}{Theorem}

\newtheorem{lemma}[thm]{Lemma}

\newtheorem{defn}[thm]{Definition}

\newenvironment{proof}{\noindent {\bf Proof --}}{\hfill$\square$ \vspace{3mm}\endtrivlist}

\catcode `\@=11 \@addtoreset{equation}{section}
\catcode `\@=12

\begin{document}

\thispagestyle{empty}
\thispagestyle{empty}
\vspace*{2cm}
\vspace{1cm}
\noindent Publisher: Springer Nature \\
Journal: Z. Angew. Math. Phys.
(2022) 73:119\\
Copyright \textcopyright 2022, The Author(s)\\
Publisher version: https://doi.org/10.1007/s00033-022-01759-z\\

\begin{center}
{\Large \bf Bi-coherent states as generalized eigenstates of the position and the momentum operators}   \vspace{2cm}\\

\vspace*{-1cm}
{\large F. Bagarello}\\
  Dipartimento di Ingegneria,
Universit\`a di Palermo,\\ I-90128  Palermo, Italy\\
and I.N.F.N., Sezione di Napoli\\
e-mail: fabio.bagarello@unipa.it\\

\vspace{3mm}

{\large F. Gargano}\\
Dipartimento di Ingegneria,
Universit\`a di Palermo,\\ I-90128  Palermo, Italy\\
e-mail: francesco.gargano@unipa.it\\
\end{center}

\vspace*{0cm}

\begin{abstract}
\noindent In this paper we show that the position and the derivative operators, $\hat q$ and $\hat D$, can be treated as ladder operators connecting the various vectors of two biorthonormal families, $\F_\varphi$ and $\F_\psi$. In particular, the vectors in $\F_\varphi$ are essentially monomials in $x$, $x^k$, while those in $\F_\psi$ are weak derivatives of the Dirac delta distribution, $\delta^{(m)}(x)$, times some normalization factor. We also show how bi-coherent states can be constructed for these $\hat q$ and $\hat D$, both as convergent series of elements of $\F_\varphi$ and $\F_\psi$, or using two different displacement-like operators acting on the two vacua of the framework. Our approach generalizes well known results for ordinary coherent states.

\end{abstract}

\vspace{2cm}


\vfill


\newpage

\section{Introduction}

The relevance of coherent states (CS) in quantum mechanics is rather well established. Since their introduction, as the {\em more classical} among the quantum states, \cite{schr}, they have been studied, refined and extended in different ways. Several applications to concrete physical systems have also been considered. A list of monographs and edited volumes on CS is the following, where the interested reader can find also many other references: \cite{klau}-\cite{Csed2}.

One of the standard ingredients when dealing with CS is a pair of ladder operators, $c$ and $c^\dagger$, which most of the times are assumed to satisfy the so-called canonical commutation relation (CCR): $[c,c^\dagger]=\1$. Here $\1$ is the identity operator on the Hilbert space $\Hil$ where $c$ and $c^\dagger$ are defined. It is useful to stress that, as it is well known, these operators are unbounded. For this reason, the CCR needs to be properly defined,  considering, for instance, its strong version $[c,c^\dagger]f(x)=f(x)$, $\forall f(x)\in\scr$, or in some other properly chosen subspace of $\Hil$. It is not really necessary to use CCR to construct CS. CS have also been constructed for fermions, see e.g. \cite{gazeaubook} and references therein. And other possible approaches also exist which still give rise to states with properties analogous to those of CS. This is, in particular, the case of {\em non-linear} CS, \cite{nlcs1,nlcs2}. Several other generalizations have been proposed during the years by many authors, and with different aims, from the vector CS to the so-called {\em Gazeau-Klauder} CS.

CS have also been considered in that slightly extended version of quantum mechanics in which the observables, and the Hamiltonian in particular, are not required to be self-adjoint. This is the case of $PT$-Quantum Mechanics, \cite{benbook}-\cite{specissue2021}, or of pseudo-Hermitian Quantum Mechanics, \cite{mosta}. We refer to \cite{fring} for some of these appearances. Another class of CS, which in our opinion is more appropriate when studied in connection with $PT$ or pseudo-Hermitian Quantum Mechanics, has been proposed in \cite{tri} and then considered in details by us in recent years. We refer to \cite{bagspringer} for a quite updated review of the results on this specific topic, with a specific view to pseudo-bosons, to their number-like operators, and to the so-called (weak) bi-coherent states: these states are (generalized) eigenstates of the two lowering operators, $a$ and $b^\dagger$, associated to the following deformation of the CCR: $[a,b]=\1$, where $a\neq b^\dagger$. The essential idea of bi-coherent states is that, since we have two lowering operators,  we could have two coherent states, $\varphi(z)$ and $\Psi(z)$. However, since $a$ and $b^\dagger$ are different, but connected, it is reasonable to imagine that $\varphi(z)$ and $\Psi(z)$ satisfy useful results only when taken {\em in pairs}, in analogy with what happens when going from orthonormal to biorthonormal bases.   The reason why we adopt here the adjective {\em generalized} for $\varphi(z)$ and $\Psi(z)$ is that, in principle, it could happen that they do not belong to $\ltwo$. In fact, this effect has been found and discussed in several papers, both in a rather general (and abstract) approach to quantum mechanics, \cite{pip,pip2}, and more recently for some concrete choices of $a$ and $b$, \cite{bagweak1}-\cite{bagweak4}. In particular, in these references, it has been shown that, when going from $[c,c^\dagger]=\1$ to $[a,b]=\1$, it might happen that $\ltwo$ is no longer the natural vector space to work with. It could 
be necessary to work with {\em compatible spaces}, \cite{bagweak4}, or even with (tempered) distributions, \cite{bagweak1}. This is due to the fact that $a$ and $b$ could be really different one from the other, as far as they satisfy  (the strong or weak form of) $[a,b]=\1$. In particular, this is true if we put $a=\hat D=\frac{d}{dx}=i\hat p$, and $b=\hat q$, the derivative and the multiplication operators. Here $\hat p$ is the momentum operator. This particular choice, first considered under this perspective in \cite{bagweak3}, will be at the basis of this paper where our main interest will be focused on the bi-coherent states associated to them. In particular,  we will show that (weak) bi-coherent states can be introduced for these operators, and how. An interesting relation with delta function of complex argument will appear as a simple consequence of our approach. We will also discuss the role of the displacement-like operators in connection with our states.

 More in details, the paper is organized as follows.  In Section \ref{sect2} we review some results on $\hat q$ and $\hat D$, considered as ladder operators on non square-integrable functions. In view of this unusual interpretation, in Section \ref{sect3} we show how these operators can be associated to specific bi-coherent states which we call {\em weak}, in view of their intrinsic distributional nature. These states are introduced via convergent series of the vectors introduced in Section \ref{sect2}. Some plots of these states are given in Section \ref{sectplots}, where we also put in evidence some differences between our states and ordinary CS. In Section \ref{sect4} we show that these states can also be introduced by acting on two different vacua with two different displacement-like operators. Our conclusions are given in Section \ref{sect5}. To keep the paper self-contained we devote Appendix A to a brief introduction to pseudo-bosons and their bi-coherent states in Hilbert spaces, while in Appendix B we discuss two interesting applications of a formula connected with the Dirac delta distribution with complex argument mentioned above.

\section{The operators $\hat q$ and $\hat D$}\label{sect2}

In the first part of this section we briefly introduce the problem, together with some of the results already deduced in \cite{bagweak3}.

Let us consider the following operators defined on $\Hil=\Lc^2(\mathbb{R})$: $\hat q f(x)=xf(x)$, $(\hat Dg)(x)=g'(x)$, the derivative of $g(x)$, for all $f(x)\in D(\hat q )=\{h(x)\in\Lc^2(\mathbb{R}): xh(x)\in \Lc^2(\mathbb{R}) \}$ and $g(x)\in D(\hat D)=\{h(x)\in\Lc^2(\mathbb{R}): h'(x)\in \Lc^2(\mathbb{R}) \}$. Of course, the set of test functions $\Sc(\mathbb{R})$ is a subset of both sets above: $\Sc(\mathbb{R})\subset D(\hat q )\cap D(\hat D)$. The adjoints of $\hat q $ and $\hat D$ in $\Hil$ are  $\hat q ^\dagger=\hat q $ and $\hat D^\dagger=-\hat D$. We have $[\hat D,x]f(x)=f(x)$, for all those $f(x)$ for which the commutator makes sense. In particular, for instance, the commutator makes sense on any $f(x)\in\scr$.  However, if we look for the vacua of $a=\hat D$ and $b=\hat q $, we easily find that, with a suitable choice of the normalizations, these are $\varphi_0(x)=1$ and $\psi_0(x)=\delta(x)$ so that neither $\varphi_0(x)$ nor $\psi_0(x)$ belong to $\Sc(\mathbb{R})$ or even to $\Lc^2(\mathbb{R})$. Nonetheless, many of the results listed in  Appendix A for pseudo-bosons can be extended to the present situation.

First of all, let us check if equation (\ref{A2}) still makes some sense. Indeed we have
\be
\varphi_n(x)=\frac{b^n}{\sqrt{n!}}\,\varphi_0(x)=\frac{x^n}{\sqrt{n!}}, \qquad \psi_n(x)=\frac{(a^\dagger)^n}{\sqrt{n!}}\,\psi_0(x)=\frac{(-1)^n}{\sqrt{n!}}\,\delta^{(n)}(x),
\label{31}\en
for all $n=0,1,2,3,\ldots$. Here $\delta^{(n)}(x)$ is the n-th weak derivative of the Dirac delta function. We see that $\varphi_n(x), \psi_n(x)\in \Sc'(\mathbb{R})$, the set of the tempered distributions, \cite{gel}, that is the set of  the continuous linear functionals on $\Sc(\mathbb{R})$. This suggests to consider $a^\dagger$ and $b$ as linear operators acting on $\Sc'(\mathbb{R})$. For this reason, in \cite{bagweak3} we have considered the (extended) action of $\hat q $ and $\hat D$ to $\Sc'(\mathbb{R})$. This was possible also because $a, b, a^\dagger$ and $b^\dagger$ all map $\Sc'(\mathbb{R})$ into itself. Hence  the following (weak) pseudo-bosonic commutation relation makes sense:
\be[a,b]\varphi(x)=\varphi(x),
\label{32}\en
for all $\varphi(x)\in\Sc'(\mathbb{R})$. In  \cite{bagweak3} the following ladder and eigenvalue equations have been deduced for the elements of $\F_{\varphi}=\{\varphi_n(x)\}$ and $\F_{\psi}=\{\psi_n(x)\}$:
\be
b\,\varphi_k(x)=\sqrt{k+1}\,\varphi_{k+1}(x), \qquad \qquad a^\dagger\psi_k(x)=\sqrt{k+1}\,\psi_{k+1}(x),
\label{33}\en
$k=0,1,2,3,\ldots$, and
\be
a\varphi_k(x)=\sqrt{k}\,\varphi_{k-1}(x), \qquad \qquad b^\dagger\psi_k(x)=\sqrt{k}\,\psi_{k-1}(x),
\label{34}\en
$k=0,1,2,3,\ldots$, with the understanding that $a\varphi_0(x)=b^\dagger\psi_0(x)=0$. Moreover, introducing $N=ba=\hat q  \hat D$, \be N\varphi_k(x)=k\varphi_k(x), \qquad\qquad N^\dagger \psi_k(x)=k\psi_k(x),\label{34bis}\en for all $k=0,1,2,3,\ldots$.  

As discussed in Appendix A, for $\D$-PBs the families of vectors $\F_{\varphi}$ and $\F_{\psi}$ are biorthogonal, and, if the vacua are chosen to satisfy $\langle\varphi_0,\psi_0\rangle=1$, they are biorthonormal. Here, the first problem is to give a meaning to the scalar product between these vectors. In fact, it is well known that, in general, two (tempered) distributions cannot be multiplied. However, see \cite{vlad}, there are exceptions: for some particular pairs of tempered distributions one can indeed define a map which extends the scalar product in $\ltwo$. And this is in fact possible for each pair $(\varphi_k(x),\psi_m(x))$, $\forall k,m\geq0$.

First we observe that the scalar product between two {\em good } functions, for instance $f(x),g(x)\in\Sc(\mathbb{R})$, can be written in terms of a convolution between $\overline{f(x)}$ and the function $\tilde{g}(x)=g(-x)$: $\left<f,g\right>=(\overline{f}* \tilde{g})(0)$. Following \cite{vlad}, in \cite{bagweak2,bagweak3} this approach was used in a quantum mechanical settings, to extend the ordinary scalar product of $\Lc^2(\mathbb{R})$ to elements $F(x), G(x)\in\Sc'(\mathbb{R})$ as the following convolution:
\be
\left<F,G\right>=(\overline{F}* \tilde{G})(0),
\label{36}\en
whenever this convolution exists. In order to compute $\left<F,G\right>$, it is therefore necessary to compute $(\overline{F}* \tilde{G})[f]$, $f(x)\in\Sc(\mathbb{R})$, that is the action of $\overline{F}* \tilde{G}$ on the test function $f(x)$, and this can be computed by using the equality  $(\overline{F}* \tilde{G})[f]=\left<F,G*f\right>$.

We refer to \cite{bagweak3} for the details of the computation. We report here only the result, which is the following: if $F(x)=x^n$ and $G(x)=\delta^{(m)}(x)$, then
$$
(\overline{F}* \tilde{G})(x)=\left\{
\begin{array}{ll}
	0 \hspace{3.6cm} \mbox{if } m>n\\
	(-1)^nn! \hspace{2.5cm} \mbox{if } m=n\\
	(-1)^m\frac{n!}{(n-m)!}x^{n-m}\hspace{0.8cm} \mbox{if } m<n,\\
\end{array}
\right.
$$ so that $(\overline{F}* \tilde{G})(0)=(-1)^n n! \delta_{n,m}$. Hence,
\be\left<\varphi_n,\psi_m\right>=\delta_{n,m},
\label{37}\en
showing that the families $\F_{\varphi}$ and $\F_{\psi}$ are biorthonormal, in our extended sense. 

As always, it is useful to check if  $\F_{\varphi}$ and $\F_{\psi}$ give rise to some resolution of the identity, as in (\ref{A4b}), for some suitable subspace of $\Hil$. In what follows we will slightly refine the results found in \cite{bagweak3}.

We first need to compute $\left<f,\varphi_n\right>$ and $\left<\psi_n,g\right>$, for suitable functions $f(x)$ and $g(x)$. It is easy to see, using (\ref{36}), that, for all $f(x)\in\Sc(\mathbb{R})$,
\be
\langle f,\varphi_n\rangle=(\overline{f}* \tilde{\varphi_n})(0)=\int_{\mathbb{R}}\overline{f(x)}\,\varphi_n(x)\,dx.
\label{38}
\en
This is not unexpected, since it is nothing that the same result we would get working formally with $f(x)$ and $\varphi_n(x)$ as if they were both square integrable functions. Incidentally we observe that $\overline{f(x)}\,\varphi_n(x)$ is integrable for all $n\geq0$, being the product of a monomial and a function in $\Sc(\mathbb{R})$. In fact, $\overline{f(x)}\,\varphi_n(x)\in\Sc(\mathbb{R})$ as well, $\forall\,n\geq0$. Formula  (\ref{36}) also implies that, $\forall g(x)\in\Sc(\mathbb{R})$,
\be
\langle g,\psi_n\rangle=(\overline{g}* \tilde{\psi_n})(0)=\int_{\mathbb{R}}\overline{g(x)}\,\psi_n(x)\,dx=\frac{1}{\sqrt{n!}}\int_{\mathbb{R}}\overline{g^{(n)}(x)}\,\delta(x)\,dx=
\frac{1}{\sqrt{n!}}\,\overline{g^{(n)}(0)},
\label{39}
\en
$n\geq0$. In deducing this formula we have also used the definition of the weak derivative of distributions, which produces, for instance, $\langle g,\delta'\rangle=-\langle g',\delta\rangle=-g'(0)$.

Let us now introduce the set ${\cal A}(\mathbb{R})$ of all those functions $f(x)$ which admit Taylor expansion  $\forall x\in\mathbb{R}$, $f(x)=\sum_{n=0}^{\infty}\,\frac{1}{n!}f^{(n)}(0)x^n$. Sometimes in the literature the elements of ${\cal A}(\mathbb{R})$ are called {\em real analytic} functions, \cite{krantz}. Then we introduce
\be
\Sc_{\cal A}(\mathbb{R})=\Sc(\mathbb{R})\cap {\cal A}(\mathbb{R}),
\label{310}\en
i.e. the set of all the functions in $\Sc(\mathbb{R})$ which are also real analytic.

\vspace{2mm}

{\bf Remark:--} It may be not so evident that the set $\Sc(\mathbb{R})$ contains also functions which are not real analytic. However
$$
p(x)=\left\{
\begin{array}{ll}
	e^{-\left(x^2+\frac{1}{x^2}\right)}, \hspace{4.9cm} \mbox{if } x\neq0\\
	0, \hspace{6.4cm} \mbox{if } x=0,\\
\end{array}
\right.
$$ 
is such a function. In fact, it is $C^\infty$ and goes to zero, together with all its derivatives, faster than any inverse power of $x$. However its $k$-th derivative in $x=0$, $p^{(k)}(0)$, is zero for all $k\geq0$, so that $p(x)$ cannot be expanded in $x=0$.

\vspace{2mm}

Before going on, we need to prove the following simple Lemma, which will be used to prove Theorem \ref{theoremQB} below:

\begin{lemma}\label{lemma1}
	Let $\{s_N(x), \,N\in\mathbb{N}\}$ be a sequence of complex-valued functions uniformly convergent to $s(x)$ in $\mathbb{R}$, and let $F(x)\in\Lc^1(\mathbb{R})$ be a given function. Suppose that $\sigma_N(x)=s_N(x)F(x), \sigma(x)=s(x)F(x)\in\Lc^1(\mathbb{R})$, $\forall N\in\mathbb{N}$. Then
	$$
	\lim_{N\rightarrow\infty}\int_{\mathbb{R}}\sigma_N(x)\,dx=\int_{\mathbb{R}}\sigma(x)\,dx.
	$$
\end{lemma}

\begin{proof}
	The uniform convergence of $s_N(x)$ to $s(x)$ implies that $\forall\epsilon>0$ $\exists N_\epsilon>0$ such that, $\forall N>N_\epsilon$, $|s_N(x)-s(x)|\leq \epsilon$, $\forall x\in\mathbb{R}$. Therefore, for all such $N$'s, $|\sigma_N(x)-\sigma(x)|\leq \epsilon|F(x)|$, $\forall x\in\mathbb{R}$. Hence
	$$
	\left|\int_{\mathbb{R}}\sigma_N(x)\,dx-\int_{\mathbb{R}}\sigma(x)\,dx\right|\leq \int_{\mathbb{R}}|\sigma_N(x)-\sigma(x)|\,dx\leq \epsilon\int_{\mathbb{R}}|F(x)|\,dx,
	$$
	which can be made as small as we want.
\end{proof}

Slightly modifying and refining what proved in \cite{bagweak3}, we now deduce the following result:
\begin{thm}\label{theoremQB}
	$(\F_\varphi,\F_{\psi})$ are $\Sc_{\cal A}(\mathbb{R})$-quasi bases.
\end{thm}

\begin{proof}
Let $f(x),g(x)\in\Sc_{\cal A}(\mathbb{R})$. Using (\ref{38}) and (\ref{39}) we have
$$
\sum_{n=0}^{\infty}\langle f,\psi_n\rangle\langle \varphi_n,g\rangle=\sum_{n=0}^{\infty}\frac{1}{\sqrt{n!}}\overline{f^{(n)}(0)}\,\int_{\mathbb{R}}\frac{x^n}{\sqrt{n!}}\,g(x)\,dx=\int_{\mathbb{R}}\overline{\sum_{n=0}^{\infty}\frac{1}{n!}f^{(n)}(0)x^n}\,g(x)\,dx=\langle f,g\rangle,
$$
using Lemma \ref{lemma1} putting $\sigma_N(x)=\overline{\sum_{n=0}^{N}\frac{1}{n!}f^{(n)}(0)x^n}\,g(x)$, $\sigma(x)=\overline{f(x)}\,g(x)$, which are both in $\Sc(\mathbb{R})$ and, therefore in $\Lc^1(\mathbb{R})$, and recalling that $f(x)\in{\cal A}(\mathbb{R})$.

In a similar way one can show that
$$
\sum_{n=0}^{\infty}\langle f,\varphi_n\rangle\langle \psi_n,g\rangle=\langle f,g\rangle,
$$
which concludes the proof.

\end{proof}

Then we can say that the operators $a=\hat D$ and $b=\hat q $ are {\em weakly pseudo-bosonic} in the sense that they and their Hermitian conjugates act on two different vacua producing two sets of distributions which are mutually orthogonal, in an extended sense, and produce a resolution of the identity on the set $\Sc_{\cal A}(\mathbb{R})$. The role of $\F_{\varphi}$ and $\F_{\psi}$ in connection with bi-coherent states will be discussed in the next section.

\section{Bi-coherent states}\label{sect3}

In Appendix A we briefly discuss how  pseudo-bosonic operators on some Hilbert space $\Hil$ can be used to construct two power series in $z\in\mathbb{C}$ which are both convergent in all the complex plane and which have some interesting properties similar to those of ordinary CS. In particular, they are eigenstates of the two pseudo-bosonic annihilation operators and produce a resolution of the identity on some dense subspace $\G$ of $\Hil$, see formulas (\ref{25}) and (\ref{27}). Here we want to show that similar results can be deduced also in our settings, where neither $\varphi_n(x)$ nor $\psi_n(x)$ can satisfy any bound like those in (\ref{22}), since they are not square integrable functions.

We start introducing the set, already considered in \cite{baginbagbook},
$$
\G=\left\{f(x)\in\Sc(\mathbb{R}): \, e^{kx}f(x)\in\Sc(\mathbb{R}), \, \forall k\in\mathbb{C}\right\}.
$$
This set is dense in $\ltwo$, since it contains $D(\mathbb{R})$, the set of compactly supported $C^\infty$ functions. It is possible to check that, $\forall f(x)\in\G$, the series $\sum_{k=0}^\infty\,\frac{z^k}{\sqrt{k!}}\,\langle f,\varphi_k\rangle$ converges for all $z\in\mathbb{C}$. More explicitly, we can check that, $\forall z\in\mathbb{C}$ and $\forall f(x)\in\G$,
\be
\sum_{k=0}^\infty\,\frac{z^k}{\sqrt{k!}}\,\langle f,\varphi_k\rangle=\int_{\mathbb{R}}\overline{f(x)} e^{zx}\,dx.
\label{41}\en
Indeed we have, taken $N\in\mathbb{N}$ and using (\ref{38}),
$$
\sum_{k=0}^N\,\frac{z^k}{\sqrt{k!}}\,\langle f,\varphi_k\rangle=\int_{\mathbb{R}}\overline{f(x)}\,\sum_{k=0}^N\frac{(zx)^k}{k!}\,dx \longrightarrow \int_{\mathbb{R}}\overline{f(x)}\,\sum_{k=0}^\infty\frac{(zx)^k}{k!}\,dx,
$$
when $N\rightarrow\infty$, so that (\ref{41}) follows. The limit $N\rightarrow\infty$ can be moved inside the integral because of Lemma \ref{lemma1}, identifying $\sigma_N(x)$ with $\overline{f(x)}\,\sum_{k=0}^N\frac{(zx)^k}{k!}$ and $\sigma(x)$ with $\overline{f(x)} e^{zx}$, which are both in $\Lc^1(\mathbb{R})$ because of the properties of the functions in  $\G$.

As it is done in \cite{bagweak4} and in \cite{bagspringer}, this suggests us to define a functional $F_\varphi$ on $\G$ as follows:
\be
F_\varphi[f](z,\overline z)=e^{-\frac{|z|^2}{2}}\sum_{k=0}^\infty\,\frac{\overline z^k}{\sqrt{k!}}\,\langle \varphi_k,f\rangle=e^{-\frac{|z|^2}{2}}\int_{\mathbb{R}} e^{\overline z x}\,f(x)\,dx,
\label{42}\en
which in turns suggests to define the function $\varphi(z;x)=e^{-\frac{|z|^2}{2}}e^{zx}$, so that we can also write
\be
F_\varphi[f](z,\overline z)=\int_{\mathbb{R}} \overline{\varphi(z;x)}\,f(x)\,dx=\langle \varphi(z),f\rangle.
\label{43}\en

{\bf Remark:--} It might be useful to notice that, even if $\varphi(z;x)$ depends on both $z$ and $x$, the dependence on $x$ does not appear in $\langle \varphi(z),f\rangle$. The reason is obvious: $x$ is integrated out, here, so that the scalar product does not depend on $x$. The same notation will be adopted in the rest of the paper, for similar quantities. 

\vspace{2mm}

The same analysis, but with some difference, can be repeated for the companion series of $\sum_{k=0}^\infty\,\frac{z^k}{\sqrt{k!}}\,\langle f,\varphi_k\rangle$, i.e. for $\sum_{k=0}^\infty\,\frac{z^k}{\sqrt{k!}}\,\langle f,\psi_k\rangle$. We start recalling that, see (\ref{39}), $\langle g,\psi_n\rangle=
\frac{1}{\sqrt{n!}}\overline{g^{(n)}(0)}$ for all $g(x)\in\Sc(\mathbb{R})$. This implies, in particular, that
\be
\sum_{k=0}^\infty\,\frac{z^k}{\sqrt{k!}}\,\langle f,\psi_k\rangle=\sum_{k=0}^\infty\,\frac{z^k}{k!}\,\overline{f^{(k)}(0)}=\overline{\sum_{k=0}^\infty\,\frac{\overline z^k}{k!}\,f^{(k)}(0)}=\overline{f(\overline z)}
\label{44}\en
for all functions in $\Sc_{\cal A}(\mathbb{R})$. Here $f(z)=\sum_{k=0}^\infty\,\frac{z^k}{k!}\,{f^{(k)}(0)}$, which is clearly convergent $\forall z\in\mathbb{C}$, because of the definition of ${\cal A}(\mathbb{R})$. As in (\ref{42}), we define a linear functional $F_\psi$ on $\Sc_{\cal A}(\mathbb{R})$ and its related {\em representation} $\psi(z;x)$, as follows
\be
F_\psi[g](z,\overline z)=e^{-\frac{|z|^2}{2}}\sum_{k=0}^\infty\,\frac{\overline z^k}{\sqrt{k!}}\,\langle \psi_k,g\rangle=e^{-\frac{|z|^2}{2}}g(\overline z)=\int_{\mathbb{R}} \overline{\psi(z;x)}\,g(x)\,dx=\langle \psi(z),g\rangle,
\label{45}\en
$\forall g(x)\in\Sc_{\cal A}(\mathbb{R})$. This could be formally rewritten as
\be
\psi(z;x)=e^{-\frac{|z|^2}{2}}\delta(x- z),
\label{46}\en
where the Dirac delta distribution with complex argument appears. We refer to \cite{complexdelta1,complexdelta2,complexdelta3} for some results on this specific topic. In Section \ref{sectplots} we plots formulas (\ref{43}) and (\ref{45}) for some specific choice of $f(x)$ and $g(x)$, and we compare the results also with the plot of ordinary CS. These plots will suggest an interesting interpretation for these states.

Using now the same steps as for ordinary CS we can check that $\varphi(z;x)$ and $\psi(z;x)$ satisfy the following resolution of the identity:
\be
\langle f,g\rangle=\int_{\mathbb{C}}\frac{d^2z}{\pi}\langle f,\varphi(z)\rangle\langle \psi(z),g\rangle=\int_{\mathbb{C}}\frac{d^2z}{\pi}\langle f,\psi(z)\rangle\langle \varphi(z),g\rangle,
\label{47}\en
for all $f(x),g(x)\in\Sc_{\cal A}(\mathbb{R})\cap\G$. Indeed we have, for instance
$$
\int_{\mathbb{C}}\frac{d^2z}{\pi}\langle f,\varphi(z)\rangle\langle \psi(z),g\rangle=\int_{\mathbb{C}}\frac{d^2z}{\pi}e^{-|z|^2}\left(\sum_{k=0}^\infty\,\frac{z^k}{\sqrt{k!}}\,\langle f,\varphi_k\rangle\right)\left(\sum_{l=0}^\infty\,\frac{\overline z^l}{\sqrt{l!}}\,\langle \psi_l,g\rangle\right)=
$$
$$
=\frac{1}{\pi}\sum_{k,l=0}^\infty\frac{\langle f,\varphi_k\rangle\langle \psi_l,g\rangle}{\sqrt{k!\,l!}}\int_{\mathbb{C}}\,d^2z e^{-|z|^2} z^k\,\overline z^l=\langle f,g\rangle,
$$
since $\int_{\mathbb{C}}\,d^2z e^{-|z|^2} z^k\,\overline z^l=\pi\delta_{l,k}k!$. The conclusion follows from Theorem \ref{theoremQB}. Incidentally we observe also that we have moved the sums outside the integral. This is, in general, a dangerous operation. However here, as formula (\ref{48}) below confirms, this can be done.

Formula (\ref{47}), together with (\ref{43}) and (\ref{45}), allows us to write
$$
\int_{\mathbb{R}}\overline{f(x)}\,g(x)\,dx=\langle f,g\rangle=\int_{\mathbb{C}}\frac{d^2z}{\pi}\left(e^{-\frac{|z|^2}{2}}\int_{\mathbb{R}}\overline{f(x)}\,e^{zx}\,dx\right)\left(e^{-\frac{|z|^2}{2}}g(\overline z)\right),
$$
which can be rewritten, changing the order of the integration,
$$
\int_{\mathbb{R}}\overline{f(x)}\,g(x)\,dx=\int_{\mathbb{R}}\overline{f(x)}\left[\int_{\mathbb{C}}\frac{d^2z}{\pi}e^{-|z|^2}e^{zx}g(\overline z)\right]\,dx.
$$
This equality should be satisfied for all  $f(x),g(x)\in\Sc_{\cal A}(\mathbb{R})\cap\G$, which is true if the following equality holds, at least weakly on $\Sc_{\cal A}(\mathbb{R})\cap\G$:
\be
\int_{\mathbb{C}}\frac{d^2z}{\pi}e^{-|z|^2}e^{zx}g(\overline z)=g(x),
\label{48}\en
$\forall g(x)\in\Sc_{\cal A}(\mathbb{R})\cap\G$. This identity looks interesting, since it can be seen as a sort of integral representation of the Dirac delta distribution with complex argument.

In fact, this equality can be checked explicitly for many functions, not necessarily in $\Sc_{\cal A}(\mathbb{R})\cap\G$. In particular, for instance, it holds for all polynomials. In Appendix B we will check (\ref{48}) for all monomials $x^n$ and for the gaussian $e^{-x^2}$.

More at an abstract level, we can recover (\ref{48}) using the following rather general idea: since $g(x)$ belongs, in particular, to $\Sc(\mathbb{R})$, it is clear that it admits a Fourier transform $\hat g(p)$ which is still in $\Sc(\mathbb{R})$, and that $g(x)=\frac{1}{\sqrt{2\pi}}\int_{\mathbb{R}}\hat g(p)\,e^{ipx}\,dp$. If we now call $\alpha$ and $\beta$ respectively the real and the imaginary parts of $z$, $z=\alpha+i\beta$, we can write
$$
g(\overline z)=\frac{1}{\sqrt{2\pi}}\int_{\mathbb{R}}\hat g(p)\,e^{ip\overline z}\,dp=\frac{1}{\sqrt{2\pi}}\int_{\mathbb{R}}\hat g(p)e^{\beta p}\,e^{ip\alpha}\,dp,
$$
which is surely well defined if, for instance, $\hat g(p)\in\G$ since, when this is true, then $\hat g(p)e^{\beta p}\in\Sc(\mathbb{R})$ by definition. Hence $g(\overline z)$ can be seen as the inverse Fourier transform of $\hat g(p)e^{\beta p}$, which exists. 

\vspace{2mm}

{\bf Remarks:--} (1) Notice that requiring that $g(x)\in\G$ does not necessarily imply that $\hat g(p)\in\G$. It only implies that $\hat g(p+ik)\in \Sc(\mathbb{R})$ for all fixed $k\in\mathbb{C}$. But this formula is not particularly useful for us, here. 

(2) However, there are important examples of functions in $\G$ whose Fourier transforms are still in $\G$. This is the case, for instance, of all the Hermite functions $e_n(x)=\frac{1}{\sqrt{2^n\,n!\,\sqrt{\pi}}}\,H_n(x)e^{-x^2/2}$. Indeed, they all belong to $\G$, of course. Moreover, since their Fourier transforms $\hat e_n(p)$'s coincide, a part some inessential factor (and a rename of the variable), with $e_n(x)$ themselves, \cite{titch},  $\hat e_n(p)\in\G$ as well.

\vspace{2mm}

Going back to the left-hand side of (\ref{48}) we have, after some rearrangement,
$$
\int_{\mathbb{C}}\frac{d^2z}{\pi}e^{-|z|^2}e^{zx}g(\overline z)=\frac{1}{\pi\sqrt{2\pi}}\int_{\mathbb{R}}dp\,\hat g(p)\int_{\mathbb{R}}d\alpha e^{-\alpha^2}e^{\alpha(x+ip)}\int_{\mathbb{R}}d\beta e^{-\beta^2}e^{\beta(p+ix)}=
$$
$$
=\frac{1}{\sqrt{2\pi}}\int_{\mathbb{R}}dp\,\hat g(p)e^{ipx}=g(x),
$$
as we had to check. Here we have used the gaussian integrals $\int_{\mathbb{R}}d\alpha e^{-\alpha^2}e^{\alpha(x+ip)}=\sqrt{\pi}e^{(x+ip)^2/4}$ and $\int_{\mathbb{R}}d\beta e^{-\beta^2}e^{\beta(p+ix)}=\sqrt{\pi}e^{(p+ix)^2/4}$.

\vspace{2mm}

{\bf Remark:--} We have deduced formula (\ref{48}) from (\ref{47}). Using the results in Appendix B, or the explicit check proposed here, we could reverse the procedure, and use (\ref{48}) to check (\ref{47}).

\vspace{2mm}

The vectors $\varphi(z;x)$ and $\psi(z;x)$ are also (weak) eigenstates of $a$ and $b^\dagger$, as any CS is expected to be. Indeed we can prove that $\forall f(x)\in\G$ and $\forall g(x)\in\Sc_{\cal A}(\mathbb{R})$, we have
\be
\langle f,a\varphi(z)\rangle=z\langle f,\varphi(z)\rangle, \qquad \langle g,b^\dagger\psi(z)\rangle=z\langle g,\psi(z)\rangle,
\label{49}\en
which are our weak versions of the eigenvalue equations for CS. The evident asymmetry in these formulas could be removed by working on a common set, $\Sc_{\cal A}(\mathbb{R})\cap\G$, but we prefer to keep this more general version of formulas (\ref{49}).

First we observe that $\G$ is closed under derivation. Indeed, if $f(x)\in\G$, then it is possible to check that $f'(x)\in\G$ as well. The proof is easy and will not be given here. Now we have, using (\ref{42}),
$$
\langle f,a\varphi(z)\rangle=-\langle f',\varphi(z)\rangle=-e^{-\frac{|z|^2}{2}}\sum_{k=0}^\infty\,\frac{z^k}{\sqrt{k!}}\,\langle f',\varphi_k\rangle=e^{-\frac{|z|^2}{2}}\sum_{k=0}^\infty\,\frac{z^k}{\sqrt{k!}}\,\langle f,\varphi_k'\rangle.
$$
Here we have used the definition of the weak derivative in the first and in the last equalities. Now, since $\varphi_0'(x)=0$ and $\varphi_k'(x)=\sqrt{k}\,\varphi_{k-1}(x)$, $k\geq1$, with standard computations we get
$$
\langle f,a\varphi(z)\rangle=e^{-\frac{|z|^2}{2}}\sum_{k=1}^\infty\,\frac{z^k}{\sqrt{k!}}\,\langle f,\sqrt{k}\,\varphi_{k-1}\rangle=ze^{-\frac{|z|^2}{2}}\sum_{l=0}^\infty\,\frac{z^l}{\sqrt{l!}}\,\langle f,\varphi_l\rangle=z\langle f,\varphi(z)\rangle,
$$
as we had to prove. In the last equality we have introduced $l=k-1$.

The proof of the second equality in (\ref{49}) works in a different way, due to the fact that the $\psi_k(x)$ are not functions (even if not square integrable, like the $\varphi_k(x)$'s), but {\em genuine} distributions.

The proof is based on the fact that, if $g(x)\in\Sc_{\cal A}(\mathbb{R})$, then $xg(x)\in\Sc_{\cal A}(\mathbb{R})$ as well. This is easy to check. Now, using (\ref{45}), we have
$$
\langle g(x),b^\dagger\psi(z)\rangle=\langle xg(x),\psi(z)\rangle=e^{-\frac{|z|^2}{2}}\,\overline{\overline z\,g(\overline z)}=e^{-\frac{|z|^2}{2}}\,z\,\overline{g(\overline z)}=z\,\langle g(x),\psi(z)\rangle,
$$
as we had to prove.

The conclusion is that, at the price of working weakly on suitable subsets of $\ltwo$, for the operators $\hat q $ and $\hat D$ it is possible to introduce two functionals $F_\varphi$ and $F_\psi$, or equivalently two $z$-dependent vectors $\varphi(z;x)$ and $\psi(z;x)$, which share with ordinary CS some of their essential properties. It is important to stress that, in what we have done so far, our weak bi-coherent states have been introduced by two convergent series, following the same underlying idea of the states introduced in Appendix A, see (\ref{24}). In  Section \ref{sect4} we will discuss the role of displacement-like operators in connection with our states.

\section{Some plots}\label{sectplots}
In this section we show some plots of the evaluations of \eqref{43} and \eqref{45} for a particular choice of $f(x)$ and $g(x)$, and we compare these plots with those of the ordinary CS, $\Phi(z,x)=\frac{1}{\pi^{1/4}}e^{-(x-\sqrt{2}\Re\{z\})^2/2+ix\sqrt{2}\Im\{z\}}$. More specifically, we will consider $f(x)\equiv g(x)=f_{\sigma}(x)=\frac{1}{\sigma\sqrt{2\pi}}e^{-\frac{|x|^2}{2\sigma^2}}$. As shown in the plots
the results are strongly dependent on the value of $\sigma$.

\begin{figure}[!h]
	\begin{center}		
		\hspace*{-0.55cm}\subfigure[]{\includegraphics[width=4.9cm]{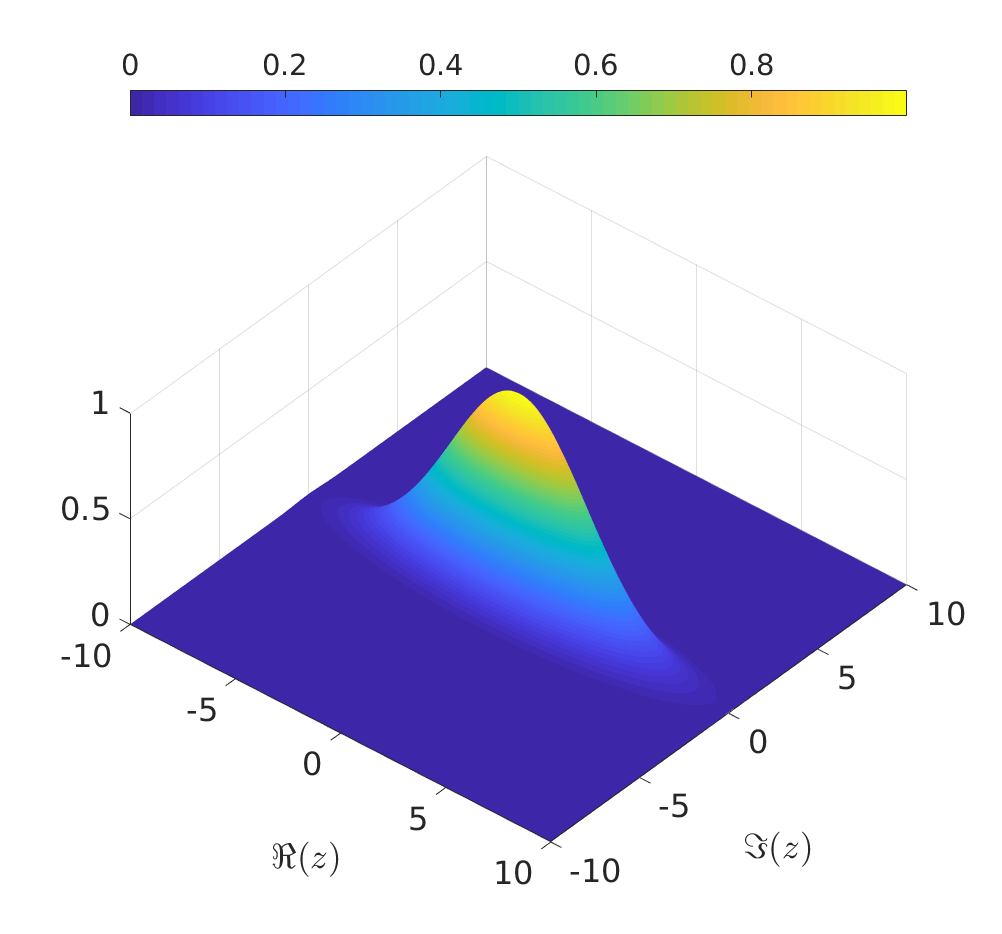}}	
		\hspace*{-0.55cm}\subfigure[]{\includegraphics[width=4.9cm]{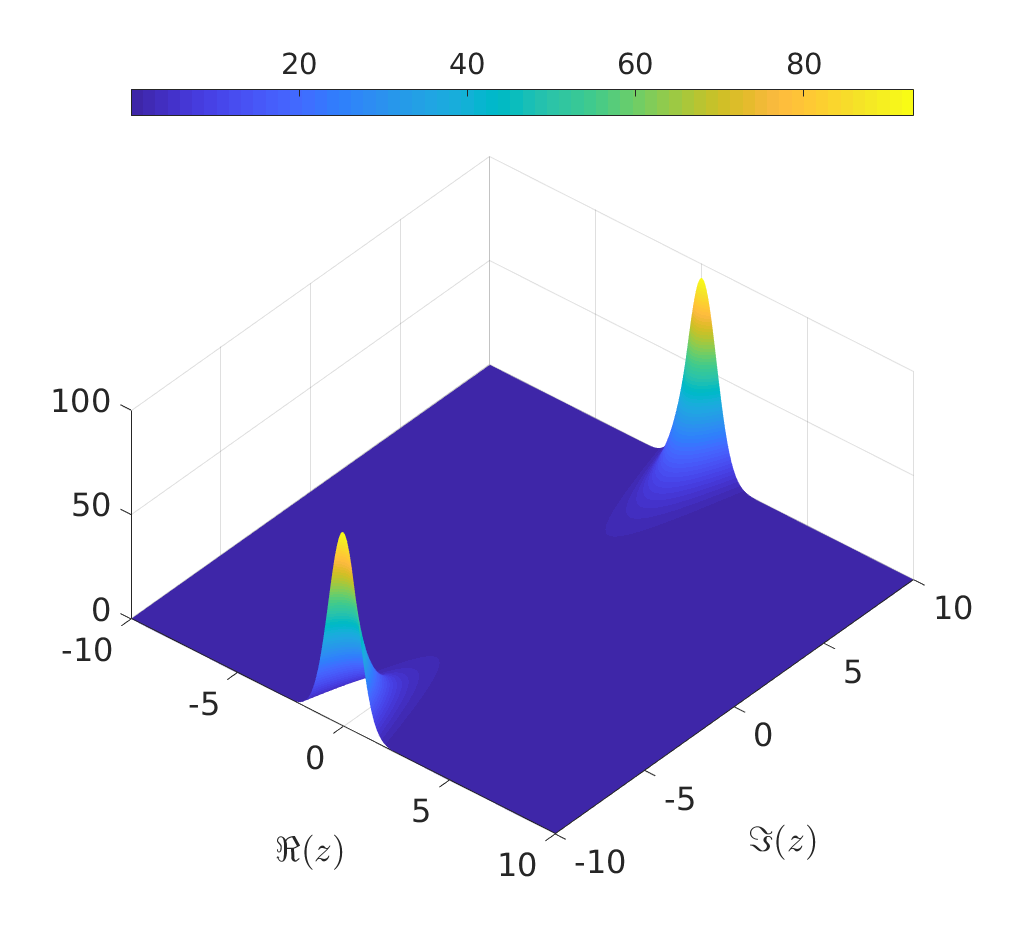}}
		\hspace*{-0.55cm}\subfigure[]{\includegraphics[width=4.9cm]{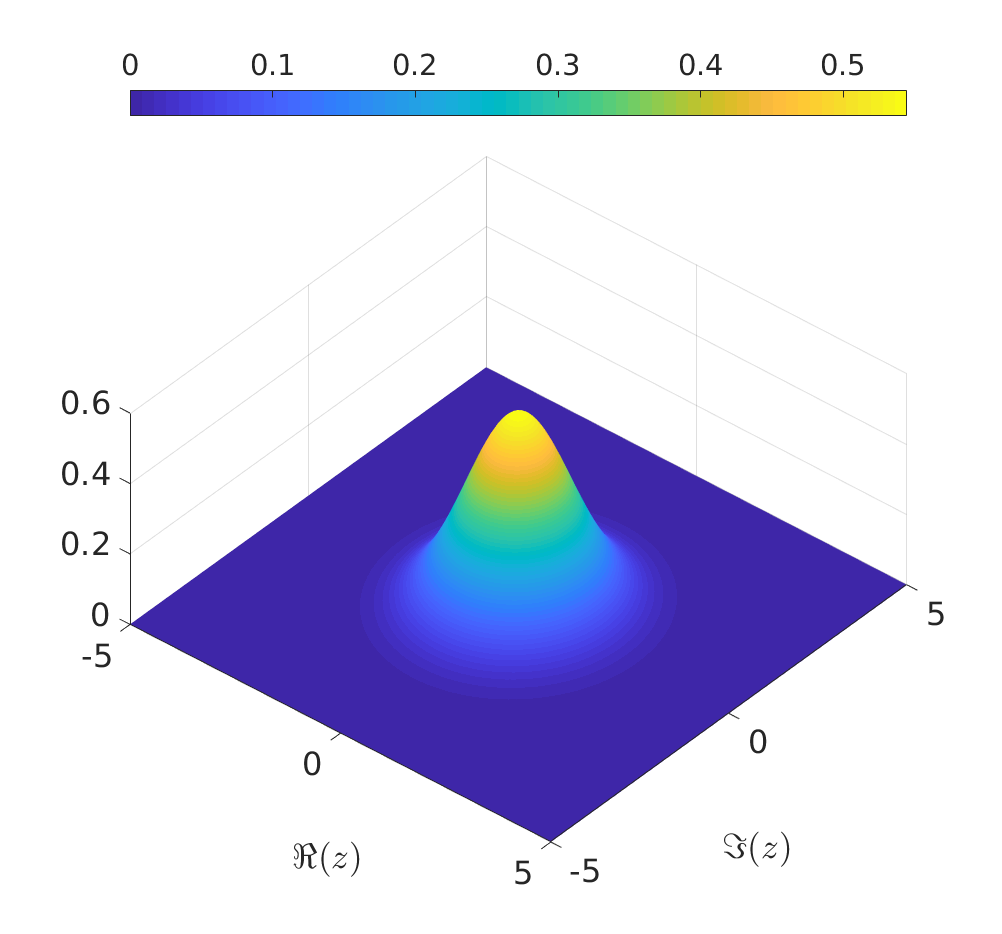}}
	\end{center}
	
	\caption{Plots of $|F_{\varphi}[f_\sigma](z, \overline z)|$ (a), $|F_{\psi}[f_\sigma](z, \overline z)|$ (b), $|F_{\Phi}[f_\sigma](z, \overline z)|$ (c) for $\sigma=0.95$.}
	\label{fig:sigam0p95}
\end{figure}

\begin{figure}[!h]
	\begin{center}		
		
		\hspace*{-0.55cm}	\subfigure[]{\includegraphics[width=4.9cm]{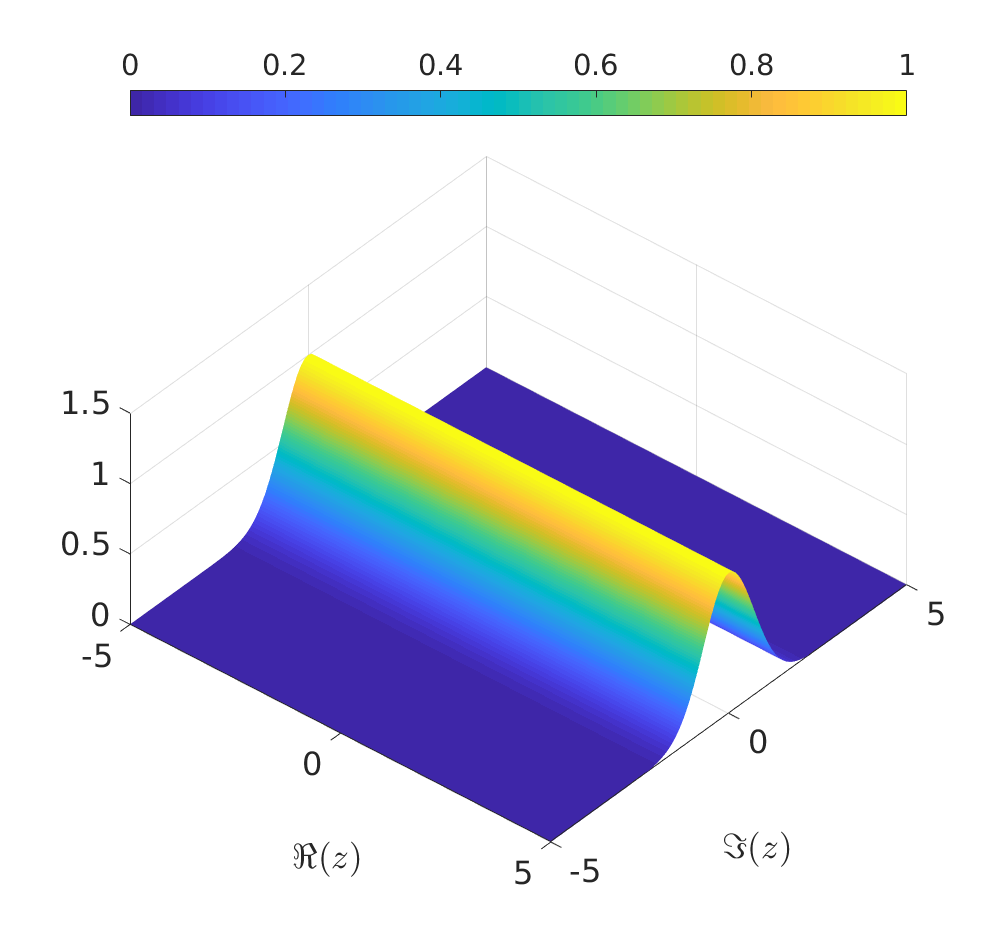}}
		\hspace*{-0.55cm}\subfigure[]{\includegraphics[width=4.9cm]{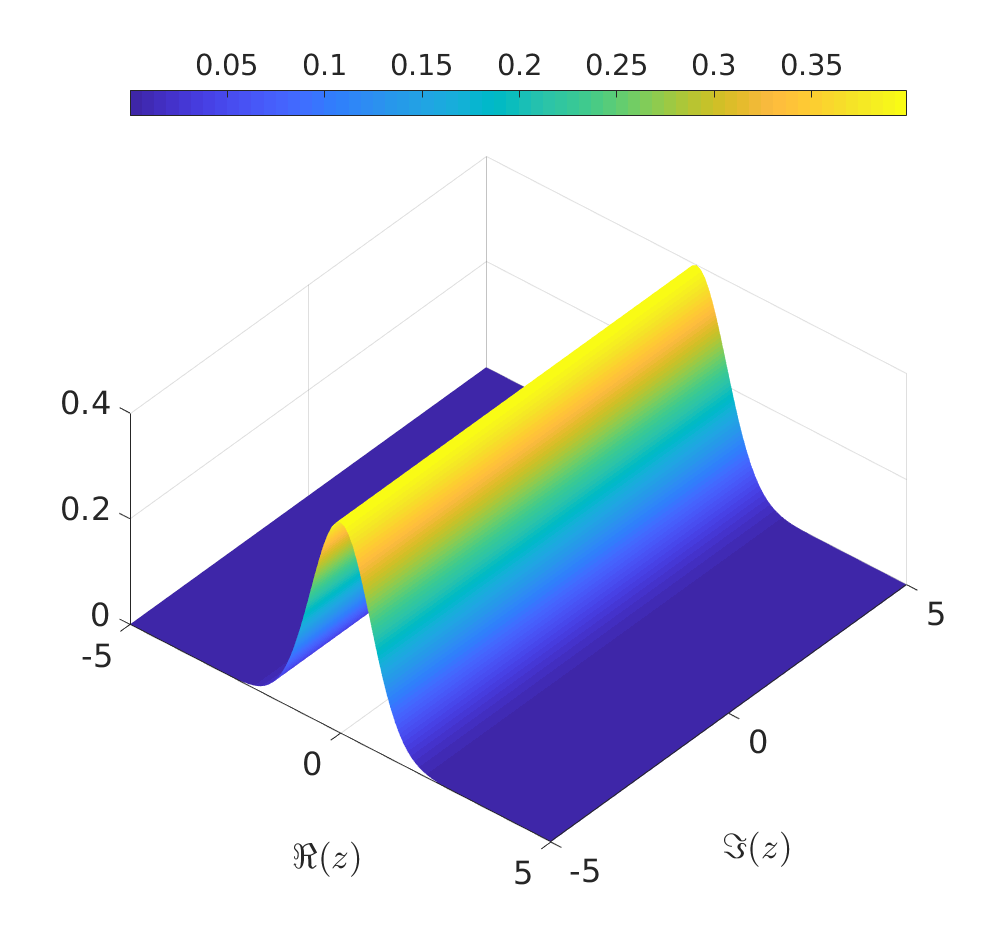}}
		\hspace*{-0.55cm}\subfigure[]{\includegraphics[width=4.9cm]{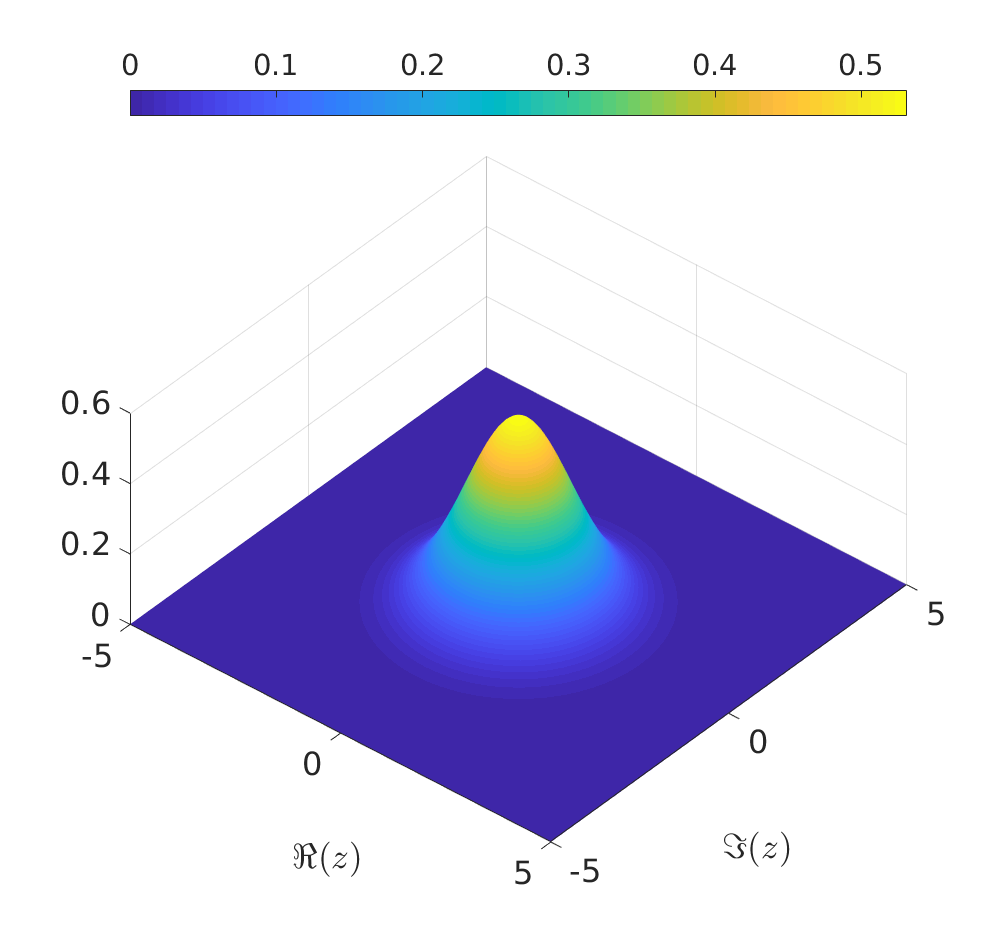}}
		
		\caption{Plots of $|F_{\varphi}[f_\sigma](z, \overline z)|$ (a), $|F_{\psi}[f_\sigma](z, \overline z)|$ (b), $|F_{\Phi}[f_\sigma](z, \overline z)|$ (c) for $\sigma=1$.}
		\label{fig:sigam1}
		
	\end{center}
\end{figure}

\begin{figure}[!h]
	\begin{center}		
		\hspace*{-0.55cm}\subfigure[]{\includegraphics[width=4.9cm]{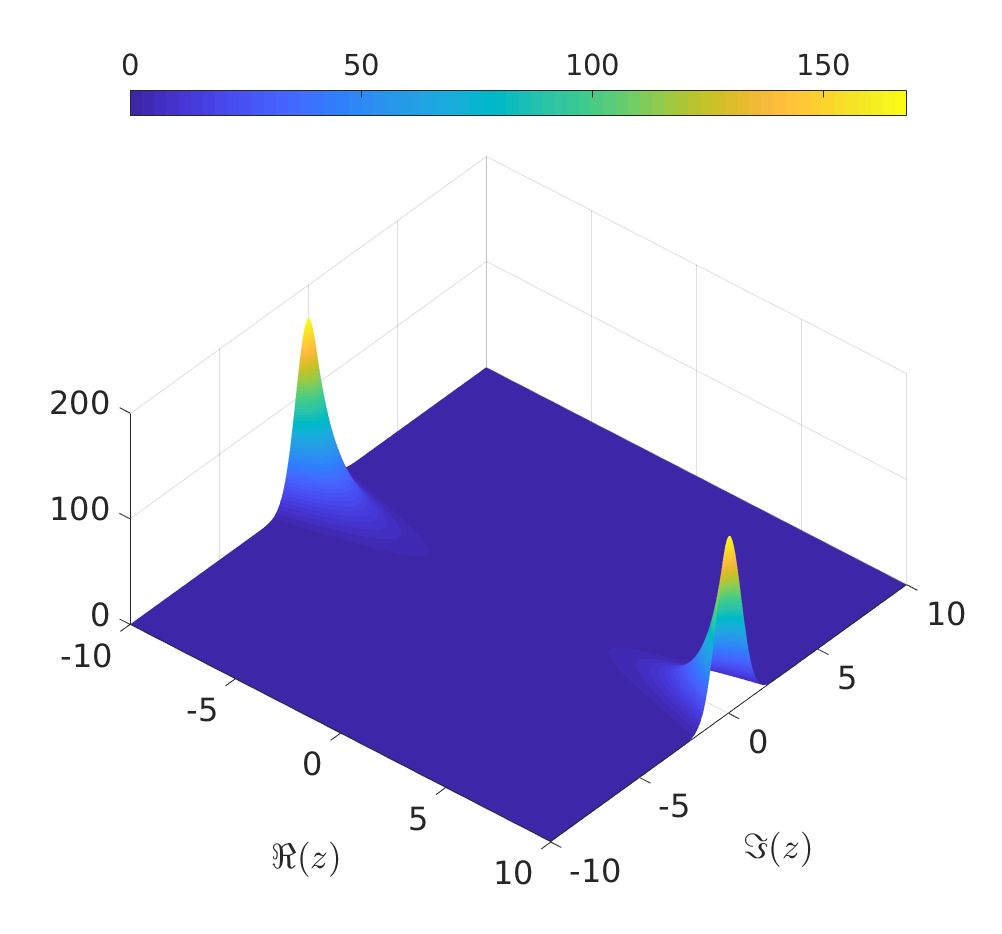}}
		\subfigure[]{\includegraphics[width=4.9cm]{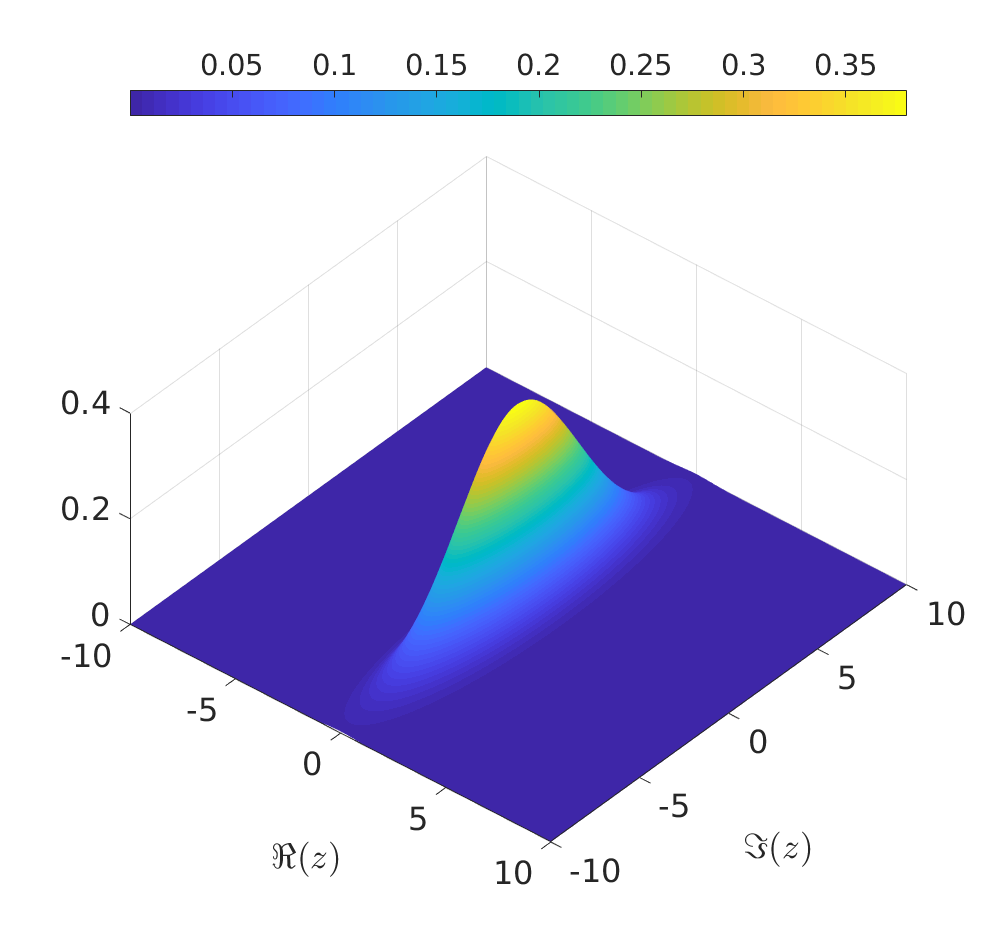}}
		\hspace*{-0.55cm}\subfigure[]{\includegraphics[width=4.9cm]{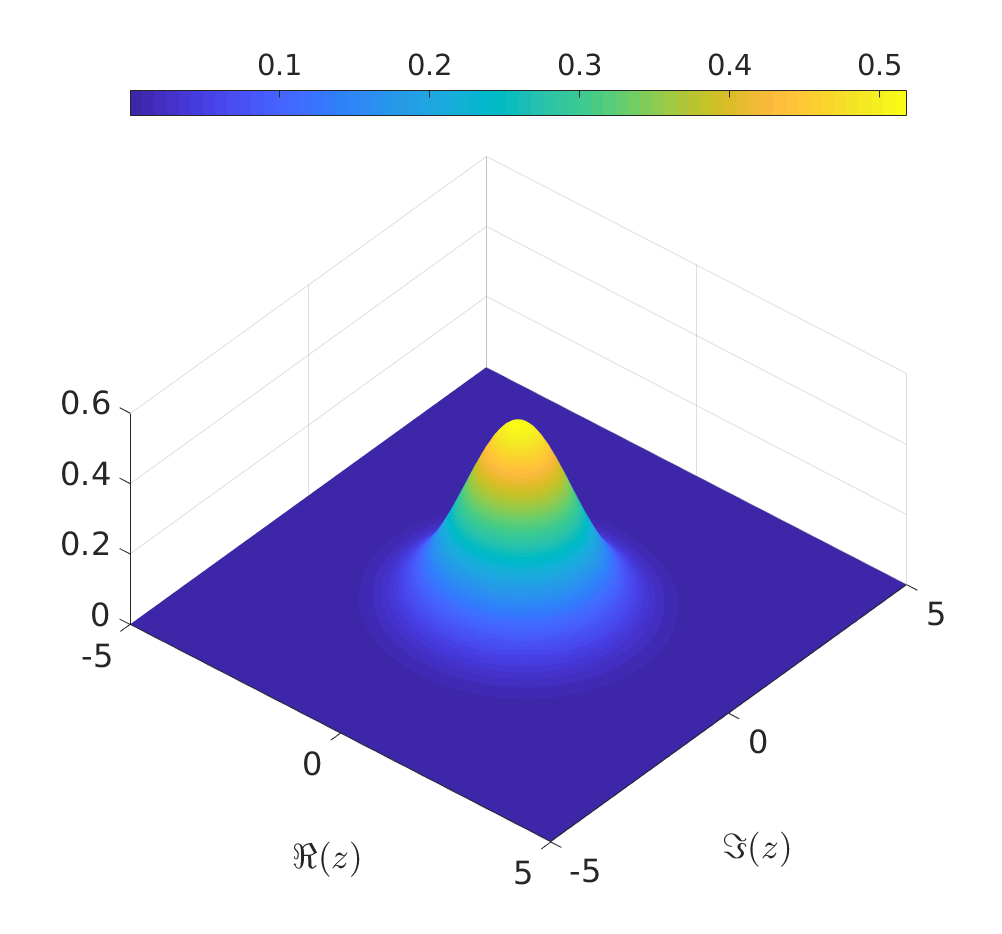}}
		
		\caption{Plots of $|F_{\varphi}[f_\sigma](z, \overline z)|$ (a), $|F_{\psi}[f_\sigma](z, \overline z)|$ (b), $|F_{\Phi}[f_\sigma](z, \overline z)|$ (c) for $\sigma=1.05$.}
		\label{fig:sigam1p05}
	\end{center}
\end{figure}

We see from Figures \ref{fig:sigam0p95}, \ref{fig:sigam1} and \ref{fig:sigam1p05}, that the behaviour of the three states can be very different. In particular, if we compare   $|F_{\varphi}[f_\sigma](z,\bar z)|$ with $|F_{\psi}[f_\sigma](z,\bar z)|$, we see that while  $|F_{\varphi}[f_\sigma](z,\bar z)|$ decays very fast along $\Im(z)$, but not along $\Re(z)$, $|F_{\psi}[f_\sigma](z,\bar z)|$ behaves exactly in the opposite way. And this is independent of the value of $\sigma$. In fact, this same behaviour is recovered also for other values of $\sigma$, other than those plotted here. On the other hand, the behaviour of the standard CS $|{F_{\Phi}[f_\sigma](z,\bar z)}|$ is the one we expect, localized in both $\Re(z)$ and in $\Im(z)$ and not so strongly dependent on $\sigma$. The three plots in each figure looks really  different. This is very close to what one of us  found also for other examples of bi-coherent states leaving in some Hilbert space, \cite{bagspringer}: the two bi-coherent states are always different one from the other, but they are connected by some sort of symmetry. Here, for instance, a part from some scaling factor, we see that  $|F_{\psi}[f_\sigma](z,\bar z)|$ for $\sigma<1$ looks like a rotated version of $|F_{\varphi}[f_\sigma](z,\bar z)|$ for $\sigma>1$. Of course, it would be interesting to understand if this is general, or strictly connected to the choices we have considered here. This question, together with other similar aspects of bi-coherent states, will be considered in a future paper.

%
%
%

\section{Displacement-like operators}\label{sect4}

We recall that, for an ordinary CS $\Phi(z)$,  we usually meets one of the following, all equivalent, expressions:
\be
\Phi(z)=D(z)e_0=e^{zc^\dagger-\overline z c}e_0=e^{-\frac{1}{2}|z|^2}\sum_{n=0}^\infty \frac{z^n}{\sqrt{n!}}\,e_n.
\label{61}\en
$D(z)=e^{zc^\dagger-\overline z c}$ is called {\em displacement operator}, and it is unitary. Here $c$ and $c^\dagger$ are the usual bosonic ladder operators satisfying $[c,c^\dagger]=\1$. In the situation considered in this paper the annihilation operator is $a=\hat D$, while the creation operator is $b=\hat q $. Hence the operator $D(z)$ should be replaced by a different operator\footnote{We indicate our displacement-like operators here and in the following with $V(z)$ and $W(z)$, even if both these operators also depend on $\overline z$. This is to simplify the notation.} which we formally write $V(z)=e^{z\hat q -\overline z \hat D}$. We could, of course, work formally with $V(z)$, for instance using the BCH-formula, \cite{gazeaubook}. But this is not what is interesting for us. We prefer to give a rigorous meaning to $V(z)$, and this will be done in the first part of this section. In particular, we will show that the vector $\varphi(z;x)$ in (\ref{43}) can be defined in complete analogy with (\ref{61}), $\varphi(z;x)=V(z)\varphi_0(x)$. In the second part of this section we will also show that the other bi-coherent state, the vector $\psi(z;x)$ in (\ref{45}), can be deduced by a second displacement-like operator $W(z)$ as in $\psi(z;x)=W(z)\psi_0(x)$, with $W(z)$ deduced out of $V(z)$ replacing  $\hat q$ and $\hat D$  by their adjoint.

\subsection{Introducing $V(z)$}\label{sect41}

Let us introduce the sequence of functions $h_n(x;z,\overline z)$ as follows
\be
h_n(x;z,\overline z)=\frac{1}{n!}\left(z\hat q -\overline z \hat D\right)^n\varphi_0(x),
\label{62}\en
$n=0,1,2,3,\ldots$. Our aim is to prove that $\sum_{n=0}^\infty h_n(x;z,\overline z)$ converges, its sum coincides with $\varphi(z;x)$, and can be used to introduce the operator $V(z)$ as follows:
$$
V(z)\varphi_0(x)=\sum_{n=0}^\infty h_n(x;z,\overline z)=\varphi(z;x).
$$

First of all we prove that $h_n(x;z,\overline z)$ can be written as follows:
\be
h_n(x;z,\overline z)=\sum_{k=0}^{\left[\frac{n}{2}\right]}\,\frac{(-1)^k}{2^k\,(n-2k)!\,k!}\,|z|^{2k}\,(xz)^{n-2k},
\label{63}\en
for all $n\geq 0$, were $\left[\frac{n}{2}\right]$ is the integer part of $\frac{n}{2}$. We begin noticing that (\ref{62}) can be rewritten as
\be
h_{n+1}(x;z,\overline z)=\frac{1}{n+1}\,\left(z\hat q -\overline z \hat D\right)h_{n}(x;z,\overline z),
\label{64}\en
$n\geq0$, with $h_0(x;z,\overline z)=\varphi_0(x)$.

In proving our claim it is convenient to distinguish between even $n$, $n=2l$, and odd $n$, $n=2l+1$, $l=0,1,2,3,\ldots$, 
We then use induction on $n$. In particular we show that (\ref{63}) returns (\ref{62}) if $n=0$ and that, when acting as in (\ref{64}), we go from $h_{2l}$ to $h_{2l+1}$, and from $h_{2l+1}$ to $h_{2(l+1)}$, covering in this way all possibilities.

If $n=0$ formula (\ref{62}) produces $h_0(x;z,\overline z)=\varphi_0(x)=1$, which is the same result we get from (\ref{63}) for $n=0$, clearly. Now, let us assume that for some $n=2l$ formula (\ref{63}) holds true:
\be
h_{2l}(x;z,\overline z)=\sum_{k=0}^{l}\,\frac{(-1)^k}{2^k\,(2l-2k)!\,k!}\,|z|^{2k}\,(xz)^{2l-2k}.
\label{65}\en 
We want to prove that, acting as in (\ref{64}), we recover the following expression for $h_{2l+1}(x;z,\overline z)$:
\be
h_{2l+1}(x;z,\overline z)=\sum_{k=0}^{l}\,\frac{(-1)^k}{2^k\,(2l+1-2k)!\,k!}\,|z|^{2k}\,(xz)^{2l+1-2k}.
\label{66}\en
The proof is long but easy. We give here only the main steps. Using (\ref{65}) we have
$$
\left(z\hat q -\overline z \hat D\right)h_{2l}(x;z,\overline z)=$$
$$=\sum_{k=0}^{l}\,\frac{(-1)^k}{2^k\,(2l-2k)!\,k!}\,|z|^{2k}\,(xz)^{2l+1-2k}-\overline z
\sum_{k=0}^{l-1}\,\frac{(-1)^k}{2^k\,(2l-2k)!\,k!}\,|z|^{2k}\,z^{2l-2k}(2l-2k)x^{2l-2k-1}=$$
$$
=\sum_{k=0}^{l}\,\frac{(-1)^k}{2^k\,(2l-2k)!\,k!}\,|z|^{2k}\,(xz)^{2l+1-2k}+\sum_{k=0}^{l-1}\,\frac{(-1)^{k+1}}{2^k\,(2l-2k-1)!\,k!}\,|z|^{2(k+1)}\,(xz)^{2l-1-2k}=
$$
$$
=\frac{(xz)^{2l+1}}{(2l)!}+(2l+1)\sum_{k=1}^{l}\,\frac{(-1)^k\,|z|^{2k}\,(xz)^{2l+1-2k}}{2^k\,(2l-2k+1)!\,k!}.
$$
In this last step we have written explicitly the contribution $k=0$ in the first sum, changed $k$ into $k+1$ in the second sum, and unified the two sums $\sum_{k=1}^{l}$ obtained in this way, simplifying the formula where possible. Hence we have
$$
\frac{1}{2l+1}\left(z\hat q -\overline z \hat D\right)h_{2l}(x;z,\overline z)=\sum_{k=0}^{l}\,\frac{(-1)^k}{2^k\,(2l+1-2k)!\,k!}\,|z|^{2k}\,(xz)^{2l+1-2k},
$$
which is $h_{2l+1}(x;z,\overline z)$ in (\ref{66}), as we had to check. 

Now we should show that
$$
\frac{1}{2l+2}\left(z\hat q -\overline z \hat D\right)h_{2l+1}(x;z,\overline z)=h_{2l+2}(x;z,\overline z),
$$
where $h_{2l+1}(x;z,\overline z)$ is given in (\ref{66}) while $h_{2l+2}(x;z,\overline z)$ can be found replacing $l$ with $l+1$ in (\ref{65}). This check is completely analogous to that described above, and will not be repeated. The conclusion is therefore that each function $h_n(x;z,\overline z)$ in (\ref{62}) can be written as in (\ref{63}).

The next step consists in computing $\sum_{n=0}^\infty h_n(x;z,\overline z)$. Indeed, if this series converges, formula (\ref{62}) suggests that this sum is what can be interpreted as $V(z)\varphi_0(x)$. In this computation it is useful to use the  identity $\sum_{n=0}^\infty\sum_{k=0}^{\left[\frac{n}{2}\right]}A_{k,n}=\sum_{n=0}^\infty\sum_{k=0}^\infty A_{k,n+2k}$, which in our computation, identifying $A_{k,n}$ with $\frac{(-1)^k}{2^k\,(n-2k)!\,k!}\,|z|^{2k}\,(xz)^{n-2k}$, returns
$$
\sum_{n=0}^\infty h_n(x;z,\overline z)=\sum_{n=0}^\infty \sum_{k=0}^{\left[\frac{n}{2}\right]}\,\frac{(-1)^k}{2^k\,(n-2k)!\,k!}\,|z|^{2k}\,(xz)^{n-2k}=\sum_{n=0}^\infty\frac{(xz)^n}{n!} \sum_{k=0}^\infty \frac{1}{k!}\left(-\frac{|z|^2}{2}\right)^k,
$$
so that
$$
\sum_{n=0}^\infty h_n(x;z,\overline z)=e^{-\frac{|z|^2}{2}}e^{zx}=\varphi(z;x),
$$
as we wanted to prove. Of course, since $\sum_{n=0}^\infty\frac{1}{n!}\left(z\hat q -\overline z \hat D\right)^n=e^{z\hat q -\overline z \hat D}=V(z)$, at least formally, we could also write 
\be
\sum_{n=0}^\infty h_n(x;z,\overline z)=e^{z\hat q -\overline z \hat D}\varphi_0(x)=V(z)\varphi_0(x)=e^{-\frac{|z|^2}{2}}e^{zx}=\varphi(z;x).
\label{67}\en
We should probably stress that this formula is not the definition of $V(z)$, since, among the other issues, it only indicates us
how $V(z)$ acts on a single state, $\varphi_0(x)$. We should also recall that $\varphi_0(x)\notin\ltwo$. Hence $\varphi_0(x)$ cannot belong to the domain of $V(z)$, strictly speaking. This is because, given an operator $T$ on some Hilbert space $\Hil$, its  domain $D(T)$ is {\em usually} meant to be a subspace of $\Hil$, \cite{reed}.

It is possible to extend the action of $V(z)$ to all monomials $\varphi_l(x)$, as we will show now. In particular, we will check that, $\forall l\geq0$, $V(z)\varphi_l(x)$ can be defined as follows:
\be
V(z)\varphi_l(x)=\sum_{n=0}^\infty\frac{1}{n!}\left(z\hat q -\overline z \hat D\right)^n\varphi_l(x)=\frac{1}{\sqrt{l!}} \, e^{-\frac{|z|^2}{2}}e^{zx}(x-\overline z)^l.
\label{68}\en
Of course, and not surprisingly, none of the functions in the right-hand side belong to $\ltwo$.   However, they are all {\em nice} functions in $x$, $z$ and $\overline z$, which makes of $V(z)$ an operator acting on a rather large set of functions in a {\em simple way}.

The proof of (\ref{68}) goes like this: we start by extending (\ref{62}). In particular we put, for all $n=0,1,2,3,\ldots$,
\be
h_n^{[l]}(x;z,\overline z)=\frac{1}{n!}\left(z\hat q -\overline z \hat D\right)^n\,x^l=\frac{1}{n!}\left(z\hat q -\overline z \hat D\right)^n\,(\sqrt{l!}\,\varphi_{l}(x)),
\label{69}\en
$l=0,1,2,3,\ldots$, and we prove that
\be
\sum_{n=0}^\infty h_n^{[l]}(x;z,\overline z)=(x-\overline z)^le^{-\frac{|z|^2}{2}}e^{zx},
\label{610}\en
for all $l\geq0$. Of course $h_n^{[0]}(x;z,\overline z)=h_n(x;z,\overline z)$, $n\geq0$, see (\ref{62}),
This, together with (\ref{67}), imply that (\ref{610}) holds for $l=0$.

Now, let us assume that (\ref{610}) holds for a given $l$. We want to check that the same equality holds also when $l$ is replaced by $l+1$. For that we will use the following formula: $$\left[\left(z\hat q -\overline z \hat D\right)^n,\hat q \right]=-n\overline z\left(z\hat q -\overline z \hat D\right)^{n-1},$$
for $n\geq1$. Of course, this commutator is zero if $n=0$. This formula can easily be derived when applied to any sufficiently regular function. In particular, it is well defined on any polynomial in $x$. We observe now that, using (\ref{69}), $h_0^{[l+1]}(x;z,\overline z)=x^{l+1}$, and
$$
h_n^{[l+1]}(x;z,\overline z)=\frac{1}{n!}\left(\left[\left(z\hat q -\overline z \hat D\right)^n,\hat q \right]+\hat q \left(z\hat q -\overline z \hat D\right)^n\right)x^l=$$
$$=-\overline z \frac{1}{(n-1)!}\left(z\hat q -\overline z \hat D\right)^{n-1}\,x^l+x \frac{1}{n!}\left(z\hat q -\overline z \hat D\right)^n\,x^l
$$
for all $n\geq1$. Hence we have
$$
\sum_{n=0}^\infty h_n^{[l+1]}(x;z,\overline z)=h_0^{[l+1]}(x;z,\overline z)+\sum_{n=1}^\infty h_n^{[l+1]}(x;z,\overline z)=$$
$$=-\overline z\sum_{n=1}^\infty h_{n-1}^{[l]}(x;z,\overline z)+x\sum_{n=0}^\infty h_{n}^{[l]}(x;z,\overline z)
=(x-\overline z)\sum_{n=0}^\infty h_{n}^{[l]}(x;z,\overline z),
$$
with a change of variable in the first sum, $n\rightarrow n-1$. Now, because of our induction assumption, formula (\ref{610}) holds for $l$. Hence we conclude that
$$
\sum_{n=0}^\infty h_n^{[l+1]}(x;z,\overline z)=(x-\overline z)^{l+1}e^{-\frac{|z|^2}{2}}e^{zx},
$$
as we had to prove.

Formula (\ref{68}) is now an easy consequence of (\ref{610}):
$$
\sum_{n=0}^\infty\frac{1}{n!}\left(z\hat q -\overline z \hat D\right)^n\varphi_l(x)=\frac{1}{\sqrt{l!}}\, \sum_{n=0}^\infty\frac{1}{n!}\left(z\hat q -\overline z \hat D\right)^n\,x^l=\frac{1}{\sqrt{l!}}\,(x-\overline z)^{l}e^{-\frac{|z|^2}{2}}e^{zx}.
$$
The output of this analysis is that, even if we are not identifying a domain for $V(z)$, we have proven that this operator can be defined on a very large set of functions. In particular, we have proven that $V(z)$ can be defined as a convergent series $\sum_{n=0}^\infty\frac{1}{n!}\left(z\hat q -\overline z \hat D\right)^n$ on any polynomial. As already observed, the fact that polynomials are not square integrable is not a problem here, since Hilbert spaces play only a minor role in the analysis considered in this paper.

\subsection{The operator $W(z)$}

The general framework of pseudo-bosons show that $a$ and $b$ are not the only ladder operators. In fact, $a^\dagger$ and $b^\dagger$ behave as ladder operators too, see (\ref{A3}). This suggests that, as widely discussed in \cite{bagspringer}, a second displacement-like operator $W(z)$ does exist, which we can formally write 
\be
W(z)=e^{za^\dagger-\overline zb^\dagger}=e^{-\overline z \hat q -z\hat D},
\label{611}\en
since $a^\dagger=-\hat D$ and $b^\dagger=\hat q $. Of course, in complete analogy with what we have seen for $V(z)$, it is possible to check that $W(z)$ can be defined on any polynomial since each series $\sum_{n=0}^\infty\frac{1}{n!}\left(-\overline z\hat q -z \hat D\right)^n x^l$ converges for all $z\in\mathbb{C}$ and for all fixed $l\geq0$. 

What is more interesting for us is to discuss the possibility to act with $W(z)$ on $\psi_0(x)$, and to see if the result is somehow related to $\psi(z;x)$ in (\ref{45}) and (\ref{46}). In other words, we want to show that $W(z)$ produces, when acting on the vacuum $\psi_0(x)$, the bi-coherent state $\psi(z;x)$. For that, we will try to understand if and how $\langle W(z)\psi_0,f\rangle$ can be defined, and if the result agrees with (\ref{45}), i.e. if
\be
\langle W(z)\psi_0,f\rangle=\int_{\mathbb{R}} \overline{\psi(z;x)}\,f(x)\,dx=e^{-\frac{|z|^2}{2}}f(\overline z).
\label{612}\en
First of all, we must clarify what  $\langle W(z)\psi_0,f\rangle$ is for us. In analogy with what we have done for $V(z)$, we will prove that, calling \be
w_n(f;z,\overline z)=\frac{1}{n!}\left\langle \left(-z\hat D-\overline z\hat q  \right)^n\psi_0,f\right\rangle,
\label{613}\en
$f(x)\in\Sc_{\cal A}(\mathbb{R})$, the series $\sum_{n=0}^\infty w_n(f;z,\overline z)$ converges for all $z\in\mathbb{C}$. Hence, in view of the formal expression (\ref{611}), we put
\be
\langle W(z)\psi_0,f\rangle=\sum_{n=0}^\infty w_n(f;z,\overline z).
\label{614}\en
As in Section \ref{sect41} we can check that $w_n(f;z,\overline z)$ can be written as the following sum:
\be
w_n(f;z,\overline z)=\sum_{k=0}^{\left[\frac{n}{2}\right]}\,\frac{(-1)^k}{2^k\,(n-2k)!\,k!}\,|z|^{2k}\,\overline z^{ n-2k}\, f^{(n-2k)}(0).
\label{615}\en
Of course, the right-hand side of this formula is well defined, since $f(x)$ is, in particular, a $C^\infty$ function. The proof of (\ref{615}) is similar to that for (\ref{63}), but with some differences. In fact, due to the need of introducing here the regularizing function $f(x)$, formula (\ref{64}) is replaced here by
\be
w_{n+1}(f;z,\overline z)=\frac{1}{n+1}w_n(\overline z\,f'-z\,x\,f;z,\overline z),
\label{616}\en
$\forall n\geq0$. This formula is a consequence of the definition of $w_n(f;z,\overline z)$ in (\ref{613}). Indeed we have, with easy computations,
$$
w_{n+1}(f;z,\overline z)=\frac{1}{n+1}\,\frac{1}{n!}\left\langle\left(-z\hat D-\overline z\hat q  \right)^n\psi_0, \left(\overline z\hat D- z\hat q  \right) f\right\rangle,
$$
from which (\ref{616}) follows. In deriving this result we have {\em moved} $(-z\hat D-\overline z\hat q )$ {\em to the right}, using in particular the definition of the weak derivative of a distribution.

It is easy to see that (\ref{613}) and (\ref{615}) return the same result, $f(0)$, when $n=0$. Next, it is possible to show that (\ref{615}) satisfies (\ref{616}) for $n$ even and for $n$ odd. In fact, the function in (\ref{615}) satisfies both
\be
w_{2l+1}(f;z,\overline z)=\frac{1}{2l+1}w_{2l}(\overline z\,f'-z\,x\,f;z,\overline z),
\label{617}\en
and
\be
w_{2l+2}(f;z,\overline z)=\frac{1}{2l+2}w_{2l+1}(\overline z\,f'-z\,x\,f;z,\overline z),
\label{618}\en
for all $l\geq0$. The proof of these statements is based on the fact that, calling $\Phi(x)=\overline z\,f'(x)-z\,x\,f(x)$, its $m$-th derivative 
$$
\Phi^{(m)}(x)=\left\{
\begin{array}{ll}
	\overline z\,f'(x)-z\,x\,f(x), \hspace{4.9cm} \mbox{if } m=0\\
	\overline z\,f^{(m+1)}(x)-m\,z\,f^{(m-1)}(x)-x\,zf^{(m)}(x), \hspace{0.8cm} \mbox{if } m\geq1,\\
\end{array}
\right.
$$
so that 
\be
\Phi^{(m)}(0)=\left\{
\begin{array}{ll}
	\overline z\,f'(0), \hspace{4.9cm} \mbox{if } m=0\\
	\overline z\,f^{(m+1)}(0)-m\,z\,f^{(m-1)}(0), \hspace{1.2cm} \mbox{if } m\geq1.\\
\end{array}
\right.
\label{619}\en
This result can be checked easily. Using now (\ref{619}), after few simple computations, we find
$$
w_{2l}(\overline z\,f'-z\,x\,f;z,\overline z)=$$
$$=\sum_{k=0}^l\frac{(-1)^k\,|z|^{2k}}{2^k\,(2l-2k)!\,k!}\,\overline z^{\, 2l-2k+1} f^{(2l-2k+1)}(0)-\sum_{k=0}^{l-1}\frac{(-1)^k\,|z|^{2k+2}}{2^k\,(2l-2k-1)!\,k!}\,\overline z^{\, 2l-2k-1} f^{(2l-2k-1)}(0)=
$$
$$
=(2l+1)\sum_{k=0}^l\frac{(-1)^k\,|z|^{2k}}{2^k\,(2l-2k+1)!\,k!}\,\overline z^{\, 2l-2k+1} \,f^{(2l-2k+1)}(0),
$$
which implies  (\ref{617}). Formula (\ref{618}) can be proved in a similar way.

Once we have proven (\ref{615}), we can compute $\sum_{n=0}^\infty w_n(f;z,\overline z)$, as in (\ref{614}). As we did for $h_n(x;z,\overline z)$, we use the identity $\sum_{n=0}^\infty\sum_{k=0}^{\left[\frac{n}{2}\right]}A_{k,n}=\sum_{n=0}^\infty\sum_{k=0}^\infty A_{k,n+2k}$, identifying now $A_{k,n}$ with $\frac{(-1)^k}{2^k\,(n-2k)!\,k!}\,|z|^{2k}\,\overline z^{ n-2k}\, f^{(n-2k)}(0)$. We find
$$
\sum_{n=0}^\infty w_n(f;z,\overline z)= \sum_{n=0}^\infty \frac{1}{n!}\,\overline z^{\, n}\,f^{(n)}(0) \sum_{k=0}^\infty \frac{1}{k!}\,\left(-\frac{|z|^2}{2}\right)^k=e^{-\frac{|z|^2}{2}}\,f(\overline z),
$$
which is formula (\ref{612}), as we had to prove. Then we can conclude that, other than $V(z)$, we can also consider the second displacement-like operator $W(z)$ giving rise, in a weak sense, the bi-coherent state $\psi(z;x)$ when acting on the vacuum $\psi_0(x)$.

\subsection{The BCH-formula}

We devote the last part of this section to check the validity of the BCH-formula for $V(z)$ and $W(z)$, when applied respectively to $\varphi_0(x)$ and $\psi_0(x)$. 

First we check that the following equalities are satisfied:
\be
e^{z\hat q -\overline z \hat D}\varphi_0(x)=e^{-\frac{|z|^2}{2}}e^{z\hat q }e^{-\overline z \hat D}\varphi_0(x)=e^{\frac{|z|^2}{2}}e^{-\overline z \hat D}e^{z\hat q }\varphi_0(x),
\label{620}\en
recalling that, as we proved in Section \ref{sect41}, $e^{z\hat q -\overline z \hat D}\varphi_0(x)=\varphi(z;x)=e^{-\frac{|z|^2}{2}}e^{zx}$.

Our check is based on the same idea used before, when we introduced $V(z)$ as a suitable convergent series. In fact, since
$$
\frac{1}{n!}(-\overline z \hat D)^n\varphi_0(x)=\delta_{n,0}\varphi_0(x),
$$
$\forall n\geq0$, the series $\sum_{n=0}^\infty\frac{1}{n!}(-\overline z\hat D)^n\varphi_0(x)$ converges clearly to $\varphi_0(x)$. This suggest, in analogy with (\ref{67}), to put $e^{-\overline z \hat D}\varphi_0(x)=\sum_{n=0}^\infty\frac{1}{n!}(-\overline z \hat D)^n\varphi_0(x)=\varphi_0(x)$. Now, recalling that $\hat q $ is the multiplication operator, we have $e^{z\hat q }e^{-\overline z \hat D}\varphi_0(x)=e^{z\hat q }\varphi_0(x)=e^{z x}\varphi_0(x)=e^{zx}$, so that the first equality in (\ref{620}) follows. To check the second, we start noticing first that, as just stated, $e^{z \hat q }\varphi_0(x)=e^{zx}$. Hence
$$
\frac{1}{n!}(-\overline z \hat D)^ne^{zx}=\frac{1}{n!}(-|z|^2)^n\,e^{zx},
$$
so that 
$$
e^{\frac{|z|^2}{2}}e^{-\overline z \hat D}e^{z\hat q }\varphi_0(x)=e^{\frac{|z|^2}{2}}\sum_{n=0}^\infty\frac{1}{n!}(-|z|^2)^n\,e^{zx}=e^{-\frac{|z|^2}{2}}\,e^{zx},
$$
which again coincides with $\varphi(z;x)$.

As for $W(z)$, we recall first that, according to (\ref{612}), $\langle W(z)\psi_0,f\rangle=e^{-\frac{|z|^2}{2}}f(\overline z)$, for all $f(x)\in \Sc_{\cal A}(\mathbb{R})$. Then we want to check if this result can be found also using the BCH-formula for $W(z)$, i.e. if we have
\be
\langle e^{-\frac{|z|^2}{2}}e^{- z \hat D}e^{-\overline z\hat q }\psi_0,f\rangle=\langle e^{\frac{|z|^2}{2}}e^{-\overline z\hat q }e^{- z \hat D}\psi_0,f\rangle=e^{-\frac{|z|^2}{2}}f(\overline z),
\label{621}\en
 for all $f(x)\in \Sc_{\cal A}(\mathbb{R})$.

 To check the first identity we use the definition of the weak derivative as follows:
 $$
 \frac{1}{n!}\langle(-z\hat D)^n \,e^{-\overline z\hat q }\psi_0,f\rangle=\frac{1}{n!}\langle \,e^{-\overline z\hat q }\psi_0,(\overline z\hat D)^nf\rangle=\frac{\overline z^{\, n}}{n!}\int_{\mathbb{R}}\overline{e^{-\overline z x}\,\psi_0(x)}\,f^{(n)}(x)\,dx=
 $$
 $$
 =\frac{\overline z^{\, n}}{n!}\int_{\mathbb{R}}\delta(x)\,e^{-zx}\,f^{(n)}(x)\,dx=\frac{\overline z^{\, n}}{n!}\,f^{(n)}(0).
 $$
 Then we have
 $$
 \langle e^{-\frac{|z|^2}{2}}e^{- z \hat D}e^{-\overline z\hat q }\psi_0,f\rangle=e^{-\frac{|z|^2}{2}}\sum_{n=0}^\infty\frac{1}{n!}\langle(-z\hat D)^n\rangle \,e^{-\overline z\hat q }\psi_0,f\rangle 
 =e^{-\frac{|z|^2}{2}}\sum_{n=0}^\infty\frac{\overline z^{\, n}}{n!}\,f^{(n)}(0)=e^{-\frac{|z|^2}{2}}f(\overline z),
 $$
 which is what we had to check. As for the second equality in (\ref{621}), we start noticing that
 $$
 \langle e^{\frac{|z|^2}{2}}e^{-\overline z\hat q }e^{- z \hat D}\psi_0,f\rangle=e^{\frac{|z|^2}{2}} \langle e^{- z \hat D}\psi_0,e^{- z x}f\rangle.
 $$
 Now, due to the fact that $e^{- z x}f(x)\in\Sc(\mathbb{R})$, we can use, as many times in this paper, the definition of the weak derivative to deduce
 $$
 \frac{1}{n!}\langle(-z\hat D)^n \psi_0,e^{- z x}f\rangle=\frac{\overline z^{\,n}}{n!}\langle \psi_0,(\hat D)^n\,e^{- z x}f\rangle=\frac{\overline z^{\,n}}{n!}\sum_{k=0}^n \left(
 \begin{array}{c}
 	n  \\
 	k  \\
 \end{array}
 \right)(-z)^{n-k}f^{(k)}(0).
 $$
 To compute the sum of all these contributions, we use now the identity $\sum_{n=0}^\infty\sum_{k=0}^{n}A_{k,n}=\sum_{n=0}^\infty\sum_{k=0}^\infty A_{k,n+k}$, and we get
  $$
  e^{\frac{|z|^2}{2}} \langle e^{- z \hat D}\psi_0,e^{- z x}f\rangle=e^{\frac{|z|^2}{2}}\sum_{n=0}^\infty\frac{(-|z|^2)^n}{n!}\sum_{k=0}^\infty\frac{\overline z^{\,k}}{k!}\,f^{(k)}(0)=e^{-\frac{|z|^2}{2}}f(\overline z),
  $$
 once again.
 
 Then we conclude that BCH formula works in the present context. Of course, this is expected but not entirely trivial since the operators we are dealing with are unbounded, and they act on distributions, rather than on usual square-integrable functions.
 
 \vspace{2mm}
 
 {\bf Remark:--} It is maybe useful to observe that here we have focused our attention only on the action of $V(z)$ and $W(z)$ on $\varphi_0(x)$ and $\psi_0(x)$, since this was enough for our purposes. Extending this result to other vectors is not an easy task, in general. 
  We refer to \cite{bagspringer} for a detailed analysis of the BCH formula in a Hilbert space settings.

\section{Conclusions}\label{sect5}

In this paper we have deduced some properties connected to the position and to the derivative (and therefore the momentum) operators arising from noticing that they can be seen as weak pseudo-bosons and, as such, they work as ladder operators on two different, but connected, sets of distributions.

In particular, after a preliminary section on these two sets, $\F_\varphi$ and $\F_\psi$, we have investigated if and how bi-coherent states can be defined for $\hat q$ and $\hat D$, and which are their properties. We have also shown that this can be done directly, by means of suitable convergent series, but also by using two different displacement-like operators, again defined as suitable convergent series.

In our knowledge, these aspects of $\hat q$ and $\hat D$ were not considered previously and open the way to several interesting mathematical problems and to possible applications to physics, and to quantum mechanics in particular.

\section*{Acknowledgements}

The authors acknowledge partial support from Palermo University and from G.N.F.M. of the INdAM. The authors thank Prof. Camillo Trapani and Dr. Federico Roccati for some fruitful discussions when preparing the paper.

\renewcommand{\theequation}{A.\arabic{equation}}

\section*{Appendix A: Few facts on pseudo-bosons and bi-coherent states}\label{appendix}

 This appendix contains a list of useful definitions and results on pseudo-bosons and on bi-coherent states in Hilbert spaces. 

\vspace{2mm}

Let $\Hil$ be a given Hilbert space with scalar product $\left<.,.\right>$ and related norm $\|.\|$. Let $a$ and $b$ be two operators
on $\Hil$, with domains $D(a)\subset \Hil$ and $D(b)\subset \Hil$ respectively, $a^\dagger$ and $b^\dagger$ their adjoint, and let $\D$ be a dense subspace of $\Hil$
such that $a^\sharp\D\subseteq\D$ and $b^\sharp\D\subseteq\D$. Here with $x^\sharp$ we indicate $x$ or $x^\dagger$. Of course, $\D\subseteq D(a^\sharp)$
and $\D\subseteq D(b^\sharp)$.

\begin{defn}\label{def21}
	The operators $(a,b)$ are $\D$-pseudo bosonic  if, for all $f\in\D$, we have
	\be
	a\,b\,f-b\,a\,f=f.
	\label{A1}\en
\end{defn}

When $b=a^\dagger$, this is simply the CCR for ordinary bosons. However, when the CCR is replaced by (\ref{A1}), the situation changes. In particular, it is useful to assume the following:

\vspace{2mm}

{\bf Assumption $\D$-pb 1.--}  there exists a non-zero $\varphi_{ 0}\in\D$ such that $a\,\varphi_{ 0}=0$.

\vspace{1mm}

{\bf Assumption $\D$-pb 2.--}  there exists a non-zero $\Psi_{ 0}\in\D$ such that $b^\dagger\,\Psi_{ 0}=0$.

\vspace{2mm}
Recalling that $\D$ is stable under the action of $b$ and $a^\dagger$, we deduce that  $\varphi_0\in D^\infty(b):=\cap_{k\geq0}D(b^k)$ and  $\Psi_0\in D^\infty(a^\dagger)$, so
that the vectors \be \varphi_n:=\frac{1}{\sqrt{n!}}\,b^n\varphi_0,\qquad \Psi_n:=\frac{1}{\sqrt{n!}}\,{a^\dagger}^n\Psi_0, \label{A2}\en
$n\geq0$, can be defined and they all belong to $\D$. Hence, they also belong to the domains of $a^\sharp$, $b^\sharp$ and $N^\sharp$, where $N=ba$.  Moreover,  the following lowering and raising relations can be easily deduced:
\be
\left\{
\begin{array}{ll}
	b\,\varphi_n=\sqrt{n+1}\varphi_{n+1}, \qquad\qquad\quad\,\, n\geq 0,\\
	a\,\varphi_0=0,\quad a\varphi_n=\sqrt{n}\,\varphi_{n-1}, \qquad\,\, n\geq 1,\\
	a^\dagger\Psi_n=\sqrt{n+1}\Psi_{n+1}, \qquad\qquad\quad\, n\geq 0,\\
	b^\dagger\Psi_0=0,\quad b^\dagger\Psi_n=\sqrt{n}\,\Psi_{n-1}, \qquad n\geq 1,\\
\end{array}
\right.
\label{A3}\en together with the eigenvalue equations $N\varphi_n=n\varphi_n$ and  $N^\dagger\Psi_n=n\Psi_n$, $n\geq0$. If  $\left<\varphi_0,\Psi_0\right>=1$, then
\be \left<\varphi_n,\Psi_m\right>=\delta_{n,m}, \label{A4}\en
for all $n, m\geq0$. Hence $\F_\Psi=\{\Psi_{ n}, \,n\geq0\}$ and
$\F_\varphi=\{\varphi_{ n}, \,n\geq0\}$ are biorthonormal. 
The analogy with ordinary bosons suggests us to consider the following:

\vspace{2mm}

{\bf Assumption $\D$-pb 3.--}  $\F_\varphi$ is a basis for $\Hil$.

\vspace{1mm}

This is equivalent to requiring that $\F_\Psi$ is a basis for $\Hil$ as well. However, several  physical models show that $\F_\varphi$ is { not} always a basis for $\Hil$, but it is still total in $\Hil$: if $f\in\Hil$ is orthogonal to $\varphi_n$, for all $n$, then $f=0$. For this reason we prefer to adopt the following weaker version of  Assumption $\D$-pb 3, \cite{baginbagbook}:

\vspace{2mm}

{\bf Assumption $\D$-pbw 3.--}  For some subspace $\G$ dense in $\Hil$, $\F_\varphi$ and $\F_\Psi$ are $\G$-quasi bases.

\vspace{2mm}
This means that, for all $f$ and $g$ in $\G$,
\be
\left<f,g\right>=\sum_{n\geq0}\left<f,\varphi_n\right>\left<\Psi_n,g\right>=\sum_{n\geq0}\left<f,\Psi_n\right>\left<\varphi_n,g\right>,
\label{A4b}
\en
which can be seen as a weak form of the resolution of the identity, restricted to $\G$.

The families $\F_\varphi$ and $\F_\Psi$ can be used to define two densely defined operators $S_\varphi$ and $S_\Psi$ via their
action respectively on  $\F_\Psi$ and $\F_\varphi$: \be
S_\varphi\Psi_{ n}=\varphi_{ n},\qquad
S_\Psi\varphi_{ n}=\Psi_{\bf n}, \label{213}\en for all $ n$. These operators play a very import role in the analysis of pseudo-bosons, since they map $\F_\varphi$ into $\F_\Psi$ and vice-versa, and define new scalar products in $\Hil$ is terms of which, for instance, the (new) adjoint of $b$ turns out to coincide with $a$. We refer to \cite{bagweak3} and \cite{baginbagbook} for more details.

According to \cite{bagspringer}, and references therein, $\F_\varphi$ and $\F_\Psi$ can also be used to construct two vectors depending on a complex variable $z$, $\varphi(z)$ and $\psi(z)$, which behave, when taken in pair, as the usual CS do, at least under some aspects. In particular, if  four strictly positive constants $A_\varphi$, $A_\Psi$, $r_\varphi$ and $r_\Psi$ exist, together with two strictly positive sequences $M_n(\varphi)$ and $M_n(\Psi)$, for which
	\be
	\lim_{n\rightarrow\infty}\frac{M_n(\varphi)}{M_{n+1}(\varphi)}=M(\varphi)>0, \qquad \lim_{n\rightarrow\infty}\frac{M_n(\Psi)}{M_{n+1}(\Psi)}=M(\Psi)>0,
	\label{21}\en
	where $M(\varphi)$ and $M(\Psi)$ could be infinity, and such that, for all $n\geq0$,
	\be
	\|\varphi_n\|\leq A_\varphi\,r_\varphi^n M_n(\varphi), \qquad \|\Psi_n\|\leq A_\Psi\,r_\Psi^n M_n(\Psi),
	\label{22}\en
	then the following series
	\be
	\varphi(z)=e^{-\frac{|z|^2}{2}}\sum_{k=0}^\infty\frac{z^k}{\sqrt{k!}}\varphi_k,\qquad \psi(z)=e^{-\frac{|z|^2}{2}}\sum_{k=0}^\infty\frac{z^k}{\sqrt{k!}}\Psi_k,
	\label{24}\en
	are all convergent in all the complex plane $\mathbb{C}$. Moreover, for all $z\in \mathbb{C}$,
	\be
	a\varphi(z)=z\varphi(z), \qquad b^\dagger \psi(z)=z\psi(z).
	\label{25}\en
	We also have
	\be
	\int_{\mathbb{C}}\left<f,\Psi(z)\right>\left<\varphi(z),g\right>\,\frac{dz}{\pi}=
	\int_{\mathbb{C}}\left<f,\varphi(z)\right>\left<\Psi(z),g\right>\,\frac{dz}{\pi}=
	\left<f,g\right>,
	\label{27}\en
	for all $f,g\in\G$.

An obvious comment is that, contrarily to what happens for ordinary coherent states, \cite{aagbook,didier,gazeaubook},  the norms of the vectors $\varphi_n$ and $\Psi_n$ need not being uniformly bounded, here. On the contrary, they can diverge with $n$, see (\ref{22}), still producing two everywhere convergent series. We refer to \cite{bagspringer} for a generalized version of this result, and for some connections of $\varphi(z)$ and $\psi(z)$ with {\em displacement-like} operators analogous to that used for ordinary coherent states, $U(z)=e^{\overline{z}\, c-z c^\dagger}$, where $[c,c^\dagger]=\1$. It is maybe useful to stress here that the vectors in (\ref{31}) do not satisfy the bounds in (\ref{22}), for any $n$. This is the reason why we have proposed in this paper a larger framework, for $\hat q$ and $\hat D$.

\renewcommand{\theequation}{B.\arabic{equation}}

\section*{Appendix B: on formula (\ref{48})}\label{appendix2}

We first check that (\ref{48}) holds for all monomials:
\be
R_n:=\int_{\mathbb{C}}\frac{d^2z}{\pi}e^{-|z|^2}e^{zx}\overline z^n=x^n,
\label{B1}\en
$n=0,1,2,3,\ldots$. We start rewriting the integral above in polar coordinates: $d^2z=rdrd\theta$. Hence
$$
R_n=\frac{1}{\pi}\int_{0}^{\infty}re^{-r^2}dr\int_{0}^{2\pi}d\theta e^{\,xre^{i\theta}}(re^{-i\theta})^n=\frac{1}{\pi}\int_{0}^{\infty}r^{n+1}e^{-r^2}dr\int_{0}^{2\pi}d\theta e^{-in\theta+rxe^{i\theta}}.
$$
But 
$
\int_{0}^{2\pi}d\theta e^{-in\theta+rxe^{i\theta}}=2\pi\frac{(rx)^n}{n!}$, and therefore
$$
R_n=2\,\frac{x^n}{n!}\int_{0}^{\infty}r^{2n+1}e^{-r^2}dr=x^n,
$$
for all allowed values of $n$.

\vspace{2mm}

To check now formula (\ref{48}) for gaussians, it is more convenient to use cartesian coordinate. For this reason we write $z=\alpha+i\beta$, so that $\int_{\mathbb{C}}d^2z=\int_{\mathbb{R}}d\alpha\int_{\mathbb{R}}d\beta$. Taking $g(x)=e^{-x^2}$ we have
$$
I=\int_{\mathbb{C}}\frac{d^2z}{\pi}e^{-|z|^2}e^{zx}g(\overline z)=\frac{1}{\pi}\int_{\mathbb{C}}d^2z e^{zx}e^{-\overline z(z+\overline z)}=\frac{1}{\pi}\int_{\mathbb{R}} d\alpha e^{-2\alpha^2+\alpha x}\int_{\mathbb{R}} d\beta e^{i\beta(x+2\alpha)},
$$
with minor computations. Now, observing that $e^{-2\alpha^2+\alpha x}$ is a test function (in the  variable $\alpha$), we can rewrite
$$
I=\frac{1}{\pi}\int_{\mathbb{R}} d\alpha e^{-2\alpha^2+\alpha x}\pi\delta\left(\alpha+\frac{x}{2}\right)=e^{-x^2},
$$
as we had to check.

\end{document}